
\input harvmac
\newcount\figno
\figno=0
\def\fig#1#2#3{
\par\begingroup\parindent=0pt\leftskip=1cm\rightskip=1cm\parindent=0pt
\baselineskip=11pt
\global\advance\figno by 1
\midinsert
\epsfxsize=#3
\centerline{\epsfbox{#2}}
\vskip 12pt
{\bf Fig. \the\figno:} #1\par
\endinsert\endgroup\par
}
\def\figlabel#1{\xdef#1{\the\figno}}
\def\encadremath#1{\vbox{\hrule\hbox{\vrule\kern8pt\vbox{\kern8pt
\hbox{$\displaystyle #1$}\kern8pt}
\kern8pt\vrule}\hrule}}

\overfullrule=0pt

%
\def\tilde{\widetilde}
\def\bar{\overline}
\def\Z{{\bf Z}}

\def\S{{\bf S}}
\def\R{{\bf R}}

\font\zfont = cmss10 

\def\bigone{\hbox{1\kern -.23em {\rm l}}}
\def\ZZ{\hbox{\zfont Z\kern-.4emZ}}

\Title{hep-th/9510135, IASSNS-HEP-95-83}
{\vbox{\centerline{BOUND STATES OF STRINGS AND $p$-BRANES}}}
\smallskip
\centerline{Edward Witten\foot{Research supported in part
by NSF  Grant PHY92-45317.}}
\smallskip
\centerline{\it School of Natural Sciences, Institute for Advanced Study}
\centerline{\it Olden Lane, Princeton, NJ 08540, USA}\bigskip

\medskip

\noindent
The recent discovery of an explicit conformal field theory
description of Type II $p$-branes makes it possible to
investigate the existence of bound states of such objects.
In particular, it is possible with reasonable  precision
to verify the prediction that the Type IIB superstring in ten dimensions
has a family of soliton and bound state
strings permuted by $SL(2,{\bf Z})$.  The space-time coordinates
enter tantalizingly in the formalism as non-commuting matrices.
\Date{October, 1995}
\newsec{Introduction}

\nref\townsend{P. Townsend, ``The Eleven-Dimensional
Super-Membrane Revisited,''   Phys. Lett. {\bf B350} (1995) 184,
hep-th/9501068.}
 \nref\witten{E. Witten,
``String Theory Dynamics In Various Dimensions,''
 Nucl.Phys. {\bf B443} (1995) 85.   hep-th/9503124.}
In many recent developments involving Type II superstrings, particles
and $p$-branes
carrying Ramond-Ramond charges have played an important role.
In many instances it is important to know about possible bound
states of such objects.  For instance, to make sense of the
strong coupling behavior of the Type IIA superstring in ten dimensions
it appears to be necessary to assume that there are Ramond-Ramond or RR
zero-branes of any (quantized) charge \refs{\townsend,\witten}.
On the other hand, to make sense of the behavior of compactified
Type II superstrings near certain conifold singularities it seems
necessary to assume that two-branes and three-branes behave under
certain circumstances as if there are no bound states \ref\strominger{A.
Strominger, ``Massless Black Holes And Conifolds In String
Theory,'' Nucl. Phys. {\bf B451} (1995) 96, hep-th/9504090.}.

\nref\hull{C. Hull and P. Townsend, ``Unity Of Superstring Dualities,''
Nucl. Phys. {\bf B438} (1995) 109, hep-th/9410167.}
One of the most interesting problems of this kind -- and the
main focus in this paper though we will also discuss other
cases -- concerns one-branes or strings in ten dimensions.
The   ten-dimensional Type IIB theory is believed to have an
$SL(2,{\bf Z})$ $S$-duality symmetry \refs{\hull,\witten}.
The two two-forms  of the theory, one from the NS-NS sector and
one from the RR sector, transform as a doublet under
$SL(2,{\bf Z})$.\foot{In general, the massless $p$-forms
of low energy supergravity theories can be assigned to     representations
of the non-compact symmetry groups even though those groups
are spontaneously broken.  Otherwise, parallel transport in the moduli
space of vacua would clash with Dirac quantization of the charges
of $p-1$-branes.}   The elementary Type IIB superstring
   is a source for
the usual $B$-field from the NS-NS sector, and not for  the RR field.
Let us describe this by saying that it has charges $(1,0)$.
$SL(2,{\bf Z})$ will map the elementary string to a string with
charges $(m,n)$ for any relatively prime pair of integers $m,n$.
The $SL(2,{\bf Z})$ prediction\foot{
This prediction is
perhaps somewhat analogous to the $S$-duality
prediction of bound states of electrons and
monopoles in $N=4$ supersymmetric
Yang-Mills theory, as checked in the two monopole sector
by Sen \ref\sen{A. Sen, ``Dyon-Monopole Bound States, Self-dual
Harmonic Forms On The Multi-Monopole Moduli Space, and $SL(2,{\bf Z})$
Invariance In String Theory,'' Phys. Lett. {\bf B329} (1994) 217,
hep-th/9402032.}}
 for the mass of these solitonic
and composite strings has been worked out in detail \ref\schwarz{J.
H. Schwarz, ``An $SL(2,{\bf Z})$ Multiplet Of Type II
Superstrings,'' hep-th/9508143,
``Superstring Dualities,'' hep-th/9509148.}.
The mass formula is  determined by the fact that the extended
strings are BPS-saturated, invariant under half of the supersymmetries
(which is also the reason that one can make sense of how the spontaneously
broken $SL(2,{\bf Z})$ acts on these strings), so it will be enough
in this paper to find a BPS-saturated string for every relatively
prime pair $m,n$ without directly computing the masses.

What makes these questions accessible for study is that an
explicit and amazingly simple description of strings (and more
general $p$-branes) carrying RR charge has recently been given
\ref\polchinski{J. Polchinski,
``Dirichlet Branes And Ramond-Ramond Charges,''  hep-th/9510017.}
in the form of $D$-strings
and $D$-branes.  The basic $D$-string has charges $(0,1)$, since
it was seen in \polchinski\ to carry  RR charge, but on symmetry
grounds
does not carry the NS-NS charge.  In section 2, we show
that the $(p,1)$ bound states exist as an almost immediate consequence
of the $D$-string structure.  To see the $(m,n)$ strings with
$n>1$ requires a more elaborate construction to which we then turn in
section 3.
In the process, two-dimensional gauge theory makes a perhaps
unexpected appearance; the existence of the BPS-saturated $(m,n)$
strings is equivalent to the existence of certain previously
unknown vacua with mass gap in two-dimensional
$N=8$  supersymmetric gauge theory!
A reasonably compelling argument for the existence of these vacua will
be presented.
We then go on in section 4 to discuss bound states
of  some of the other
 $p$-branes.

One of the most interesting       aspects of the formalism is that
the space-time coordinates normal to the $p$-brane world-sheet
-- or all of them for instantons, that is for $p=-1$ -- enter
as non-commuting matrices (in the adjoint representation of $SU(n)$
in the case of a bound state of $n$ $D$-branes).  Though
perhaps easily
motivated  given the relation of $D$-branes to Chan-Paton factors via
$T$-duality
\nref\oldpolch{J. Dai, R. G. Leigh, and J. Polchinski,
``New Connections Between String Theories,'' Mod. Phys. Lett. {\bf A4}
(1989) 2073.}
\nref\otherpolch{J. Polchinski, ``Combinatorics of Boundaries In
String Theory,'' Phys. Rev. {\bf D50} (1994) 6041, hep-th/9407031.}
\nref\horava{P. Horava, ``Strings On World-Sheet Orbifolds,''
Nucl. Phys. {\bf B327} (1989) 461, ``Background Duality Of Open
String Models,'' Phys. Lett. {\bf B321} (1989) 251.}
\nref\green{M. B. Green, ``Space-Time Duality And Dirichlet
String Theory,'' Phys. Lett. {\bf B266} (1991) 325.}
\nref\sagnotti{A. Sagnotti, ``Open Strings And Their Symmetry
Groups,'' in Cargese `87, ``Nonperturbative Quantum Field Theory,''
ed. G. Mack et. al. (Pergamon Press, 1988) p. 527.}
\refs{\oldpolch - \sagnotti}, this seems possibly
significant for the general understanding of string theory.

\newsec{The $(n,1)$ Strings}

First we recall the structure of $D$-branes, and verify that the
$D$-string has the same low-lying world-sheet excitations as the
fundamental Type IIB superstring and so can be interpreted as
the desired $(0,1)$ string.

We work in ten-dimensional Minkowski space, with a time coordinate
$x^0$ and space coordinates $x^1,\dots ,x^9$.  Ordinarily the
Type II theory has closed strings only.  The Dirichlet $p$-brane
-- with $p$ even or odd for Type IIA or Type IIB --
is simply an object whose presence modifies the allowed boundary
conditions of strings so that in addition  to the usual Neumann
boundary conditions, one may also have Dirichlet boundary conditions
with (for example) $x^{p+1}=\dots=x^9=0$.  How to compute the mass per
unit volume
of a $p$-brane was explained in \otherpolch, while the fact that
the $p$-brane couples to the appropriate RR $p+1$-form is
explained in \polchinski.

Once such  an object is introduced, in  addition to the usual
closed strings, one can have open strings whose ends are at any
values of $x^0,\dots,x^{p}$ with $x^j=0$ for $j>p$.
These strings describe the excitations of the $p$-brane.  The quantization
of these open strings is isomorphic to the conventional quantization
of an oriented open superstring.  The massless states are a vector and
a spinor making up a ten-dimensional supersymmetric Yang-Mills
multiplet with gauge group $U(1)$.
As the zero modes of $ x^j,\,j>p$ are eliminated
by the boundary conditions, the massless particles are functions
only of $x^0,\dots,x^p$.  The massless bosons $A_i(x^s)$, $i,s=0,\dots
p$ propagate as a $U(1)$ gauge boson on the $p$-brane world-surface, while
the other components $\phi_j(x^s)$ with  $j>p$, $s=0,\dots,p$,
 are scalars in the $p+1$-dimensional
sense.  Note that the vectors have conventional open-string gauge
boson vertex operators $V_A=\sum_{i=0}^p
A_i(X^s)\partial_\tau X^i$, with $\partial_\tau$ the
derivative   tangent to the world-sheet boundary, while (as in
\otherpolch) the
scalars have vertex operators of the form $V_\phi=\sum_{j>p}\phi_j(X^s)
\partial_\sigma X^j$ with $\partial_\sigma$ the normal derivative
to the boundary.   For $\phi_j=\,{\rm constant}$, the boundary integral
of $V_\phi$ is the change in the world-sheet action upon adding a constant
to $X^j,\,j>p$, so the scalars can be interpreted as oscillations
in the position of the $p+1$-brane.  This interpretation is
developed in some detail in \ref\leigh{R. Leigh,
``Dirac-Born-Infeld Action From Dirichlet Sigma Model,''
Mod. Phys. Lett. {\bf A4} (1989) 2767.}.
The theory on the $p+1$-dimensional world-volume is naturally
thought of as the ten-dimensional $U(1)$ supersymmetric gauge theory
dimensionally reduced to $p+1$ dimensions.

\def\Z{{\bf Z}}
In particular, consider the one-brane or $D$-string of the Type IIB
theory, in the conventional Lorentz background with the
RR scalar vanishing.  In this case the $U(1)$ gauge field $A_i,\,
i=0,1$ describes no propagating degrees of freedom, though it
will nonetheless play an important role latter.  The massless
bosons are the transverse oscillations of the string.  As for
the massless fermions, after the GSO projection they are a spinor
$\psi_\alpha,$ $\alpha = 1\dots 16$, with definite chirality in the
ten-dimensional sense.  The left and right-moving fermions on the
$D$-brane world-sheet are thus components of ten-dimensional
spinors of the {\it same} chirality (they are even components
of the same spinor), as is usual for Type IIB.  Thus,   the
$D$-string has, as expected, the same world-sheet structure
as the elementary Type IIB superstring, making possible its
interpretation as the $(0,1)$ $SL(2,{\bf Z})$ partner of the $(1,0)$
elementary string.

\bigskip \noindent
{\it Bound States Of Fundamental Strings With A $D$-String}

Now we begin our study of bound states by looking for the
$(m,1)$ states, that is, the bound states of $m$ fundamental
strings with a $D$-string.  To see that something special
must be at work, it is enough to consult the expected mass formula
\schwarz.  For the case in which the RR scalar vanishes, the
tension in string units of an $(m,n)$ string is
\eqn\ikko{T_{m,n}=T\sqrt{ m^2+{n^2\over\lambda^2}} }
with $T$ a constant and $\lambda$ the string coupling constant.
In particular, $T_{0,1}\sim1/\lambda$, an example of
the fact that RR charges have mass
of order $1/\lambda $ in string units.

The point about \ikko\ which is perhaps surprising at first
sight is that the binding energy of an elementary string with
a $D$-string to form a $(1,1)$ string is almost one hundred
per cent.  In fact,
an elementary string has a tension of order one,
but the difference in tension between a $(0,1)$ $D$-string
and a $(1,1)$ string is of order $\lambda$, so to first order
the tension energy of an elementary string completely disappears
when it combines with a $D$-string.

\def\R{{\bf R}}
\def\S{{\bf S}}
While this may sound surprising, it is easy to see in the $D$-string
picture why  it is so.
To make the discussion clean, compactify the $x^1$ direction,
so that we are working on $\R\times \S^1\times \R^8$, where
$\R$ is the time direction, $x^1$ is the spatial direction of
the strings we will consider,
and $\R^8$ encompasses the normal directions.
Consider first, in the absence of $D$-strings, a collection
of ordinary strings wrapping around $\S^1$.  The total string winding
number is conserved because the closed strings are not permitted
to break, or alternatively because after integrating
over $\S^1$ there is a gauge field on
the non-compact low energy world $\R\times \R^8$
-- namely the component $B_{i\,1}$ of the Neveu-Schwarz $B$ field --
whose conserved charge is the total winding number of strings.

Now suppose that there is also a $D$-string wrapping around the $\S^1$,
described as above by allowing strings to terminate at
$x^2=\dots = x^9=0$.  In the presence of such a $D$-string,
the winding number of elementary strings around the $\S^1$ is not
conserved!  In fact, let $V_W$ be the vertex operator of an elementary
string wrapping a certain number of times around the $\S^1$.  In
the presence of the $D$-string, $V_W$ has a non-zero one point function
on the disc, because the boundary of the disc, while fixed
at $x^j=0,\,j>1$, is permitted to wrap around $\S^1$ any required
number of times.

\def\F{{\cal F}}
Since there is a conserved gauge charge coupled to the
elementary string winding states, how can it be that the winding
states can disappear in the $D$-brane sector?  Obviously, the
conserved charge must be carried also by some     degrees of freedom
on the $ D$-brane world-sheet.  This in fact happens as a result
of a mechanism that has long been known
\ref\cremmer{E. Cremmer and J. Scherk, ``Spontaneous Dynamical
Breaking Of Gauge Symmetry In Dual Models,''
 Nucl. Phys. {\bf B72} (1974) 117.}
in the context of
oriented open and closed bosonic strings.
The Neveu-Schwarz $B$ field couples in bulk to the string world-sheet
$\Sigma$ by an interaction
\eqn\huggo{I_B={1\over 2}
\int_\Sigma d^2\sigma\epsilon^{\alpha\beta}\partial_\alpha
X^i\partial_\beta X^j B_{ij},}
Under a gauge transformation $B_{ij}\to B_{ij}+\partial_i\Lambda_j
-\partial_j\lambda_i$, \huggo\ is invariant if the boundary of
$\Sigma$ is empty.  If not, then $I_B$ transforms as
\eqn\kuggo{I_B\to I_B+\int_{\partial\Sigma}d\tau \,\,\Lambda_i{dX^i\over
d\tau}.}
Before concluding that the theory with the $B$ field is inconsistent
in the presence of boundaries, one must remember that the oriented
open string has a $U(1)$ gauge field $A$, encountered above,
which couples to the boundary by
\eqn\juggo{ I_A=\int_{\partial\Sigma}d\tau A_i {dX^i\over d\tau}.}
Gauge invariance is therefore restored if the gauge transformation of
$B$ is accompanied by $A\to A-\Lambda$.  The gauge-invariant
field strength is therefore not $F_{ij}=\partial_iA_j-\partial_jA_i$
but $ \F_{ij}=F_{ij}-B_{ij}$.

In the case at hand,
the $B$ kinetic energy is an integral over all space ${\cal M}$,
while
the  gauge field ``lives'' only on the $D$-brane world-sheet $\Sigma$.
The  relevant part of the effective action therefore scales  like
\eqn\pluggo{I=\int_{\cal M}d^{10}x\sqrt g {1\over 2\lambda^2}|dB|^2
              +\int_\Sigma d^2x \sqrt{g_\Sigma}{1\over 2\lambda}
{\cal F}^2}
where now ${\cal F}=\epsilon^{ij}{\cal F}_{ij}/2$ and $g,g_\Sigma$
are the metrics on ${\cal M}$ and on $\Sigma$.  The scaling
with $\lambda$ comes from the fact that the first term comes
from the two-sphere, and the second from the disc.

A gauge field in two dimensions has no propagating degrees of freedom,
and indeed the equation of motion of $A$ asserts that ${\cal F}$ is
constant.  The equation of motion for $B$ contains ${\cal F}$ as a source,
so $D$-brane states with non-zero ${\cal F}$ carry the $B$-charge;
these are the sought-for $D$-brane states carrying the charge that
in the absence of $D$-branes is carried only by string winding states.

The momentum conjugate to $A$ is $\pi = {\cal F}/\lambda$,
and as we will see momentarily $\pi$ is quantized in integer units,
so ${\cal F}=m\lambda$ for integer $m$. With this value of ${\cal F}$,
the ${\cal F}$ dependent part of the energy is of order
 of order $ m^2\lambda$ as expected from \ikko.

Quantization of $\pi$ as claimed above is equivalent to the statement
that the gauge group is $U(1)$ rather than ${\bf R}$.  The Hamiltonian
constraints of      two-dimensional gauge theory require
that the states  be invariant under infinitesimal
gauge transformations and so are linear combinations of the states
$\psi_m(A)=\exp(im\int_{{\bf S}^1} A)$.   The  state $\psi_m$
 -- which
has $\pi=m$ --  is only invariant under topologically non-trivial $U(1)$
gauge transformations if $m$ is an integer.  It must be the case
that the gauge group is $U(1) $ and $m$ is quantized, because if
$D$-strings could carry a continuum of values of the elementary
string winding number, then  annihilation of $D$-strings and
antistrings
would yield an inexistent continuum of string winding number states
in the vacuum sector.\foot{Though the principles for $p$-forms
of $p>1$ are less clear, an analogous mechanism involving compactness
of the gauge group may well cause the vacuum parameter discussed
at the end of \polchinski\ to be quantized.}

If one  wants the allowed values of $\pi$ to be not integers $m$, but
rather to be of the form $m+\theta/2\pi$ with some fixed $\theta$, then
the method to achieve that is well known \ref\coleman{S. Coleman,
``More About The Massive Schwinger Model,'' Ann. Phys. {\bf 101} (1976)
239.}.
One must add to the action a term of the form $\int_\Sigma
{\rm constant}\cdot {\cal F}$.  In our problem the ``constant''
must be the scalar field $\beta$ from the RR sector; the desired coupling
should come from a $\beta - A$ two point function on the disc.
This shift in the allowed values of the elementary string winding
number for $D$-strings appears in the general $SL(2,{\bf Z})$-invariant
formula \schwarz\ for the tension of $D$-strings.  It is analogous
to the $\theta$-dependent shift in the allowed  electric   charges of
a magnetic monopole \ref\oldwitten{E. Witten, ``Dyons Of Charge
$e\theta/2\pi$,''  Phys. Lett. {\bf 86B} (1979) 283.}.

There actually is a good   analogy between the problem we have
just analyzed and certain questions about monopoles in, say,
$N=4$ super Yang-Mills theory in four dimensions. In that theory,
the BPS formula for masses of electrons and dyons shows that
for weak coupling the rest mass of an electron nearly disappears
when it combines with a monopole.  That occurs because the classical
monopole solution is not invariant under charge rotations;
there is therefore a collective coordinate which for weak coupling
can carry charge at very little cost in energy.  For strings,
the gauge transformation that measures string winding number is
a constant mode of the gauge parameter $\Lambda$ of the $B$ field.
The equation $\delta A=-\Lambda$ shows that the $D$-string
is not invariant under this gauge transformation, so that there
is a collective coordinate -- namely $\int_{{\bf S}^1} A $ --
that can carry string winding number at very little cost in energy.

\newsec{Bound States Of $D$-Strings}

We will here describe the general setting for analyzing bound
states of $D$-branes and then apply it to $D$-strings.

First of all, to find a bound state of $n$ parallel Dirichlet $p$-branes,
one needs a description of the relevant low energy physics
that is valid when the branes are nearby.
Because of the relation of $D$-branes to Chan-Paton factors,
there is an obvious guess for what happens when the  $D$-branes
are precisely on top of each other: this should correspond to
an unbroken $U(n)$ gauge symmetry, with the effective action
being ten-dimensional $U(n)$ supersymmetric gauge theory,
dimensionally reduced to the $p+1$-dimensional world-volume of
the $D$-brane.

This ansatz can be justified by considering two parallel Dirichlet
$p$-branes,
say one at $x^j=0$, $j>p$, and one at $x^j=a^j$, $j>p$.  To find
the excitation spectrum in the presence of two $p$-branes, in addition
to closed strings, we must consider strings that start and end on
one or the other $p$-brane.  Strings  that start and end on the
same brane give a $U(1)\times U(1)$ gauge theory (with one $U(1)$
living one each $p$-brane) as we have discussed at the beginning
of section 2.  We must also consider strings starting at the first
brane and ending at the second, or vice versa.  Such strings
have $U(1)\times U(1)$ charges $(-1,1)$ or $(1,-1)$, respectively.
The ground state in this sector (as usual for supersymmetric open
strings) is a vector multiplet; as
 the string must stretch from the origin to $a=(a^{p+1},\dots,
a^9)$, the vector multiplet mass,
which is the ground state energy in this sector, is $T|a|$ with
$T$ the elementary string tension.  Thus as $a\to 0$, world-volume
vector bosons of charges $(\mp 1, \pm  1)$ becomes massless, giving
the restoration of $U(2)$  gauge symmetry on the world-volume.
Likewise, as $n$ parallel
branes become coincident, one has restoration of a $U(n)$ gauge
symmetry on the world-volume of the $p$-brane.  Filling out the
vector multiplets, the low energy theory on the world-volume
is simply the dimensional reduction of ten-dimensional supersymmetric
Yang-Mills theory with gauge group $U(n)$ from ten to $p+1$ dimensions.

To make the physical interpretation somewhat clearer, let us consider
in some detail the bosonic fields of this theory.  In addition
to the world-volume $U(n)$ gauge field $A$, there are $9-p$ scalar
fields in the adjoint representation of $U(n)$.
Changing notation slightly
from section 2, we will call them $X^j$, $j=p+1,\dots ,9$, because
of their interpretation, which will be clear momentarily, as position
coordinates of the $D$-branes.
The potential energy for the $X^j$ in the dimensional reduction
of ten-dimensional supersymmetric Yang-Mills theory is
\eqn\murf{V={T^2\over 2}\sum_{i,j=p+1}^9 \Tr\,[X^i,X^j]^2.}
Classical states  of unbroken supersymmetry
or zero energy thus have $[X^i,X^j]=0$, so the $X^i$ can
be simultaneously diagonalized, giving $ X^i={\rm diag}(a^i_{(1)},
a^i_{(2)},\dots,a^i_{(n)}).$  We interpret such a configuration as
having $n$ parallel  $p$-branes with the position of the
$\lambda^{th}$ $p$-brane, $\lambda=1,\dots, n$ being the space-time
point $a_\lambda$ with coordinates
$a^i_{(\lambda)}$.  As support for this interpretation, note
that the masses obtained by expanding \murf\ around the given
classical solution are just $T|a_\lambda-a_\mu|,$
$1\leq \lambda<\mu\leq n$, corresponding to strings with one
end at $a_\lambda$ and one at $a_\mu$.  Note that the diagonalization
of the $X^j$ is not unique; the Weyl group of $U(n)$ acts by permuting
the $a_\lambda$'s, corresponding to the fact that the $n$ $p$-branes
are identical bosons (or fermions).

In sum, then, the naive configuration space of $n$ parallel, identical
$p$-branes should be interpreted as the moduli space of classical
vacua of the supersymmetric $U(n)$ gauge theory.  When the $p$-branes
are nearby, the massive modes cannot be ignored; one must then
study the full-fledged quantum $U(n)$ gauge theory, and not
just the slow motion on the classical moduli space of vacua.
The appearance of non-commuting matrices $X^j$ which can be interpreted
as space-time coordinates to the extent that they commute is
highly intriguing for the future.  In what follows, we use this
formalism to study string and $p$-brane bound states.

\bigskip\noindent
{\it The Mass Gap}

In fact, a BPS-saturated  state of $n$ $p$-branes would correspond
simply to a supersymmetric ground state of the effective Hamiltonian,
that is to a supersymmetric vacuum of the $p+1$-dimensional supersymmetric
gauge theory.  Let us discuss just what kind of vacuum would correspond
to a {\it bound state}.

A {\it bound state} is a state that is normalizable except for the
center of mass motion.  In the problem at hand, there is no difficulty
in separating out the overall motion of the center
of mass.  A $U(n)$ gauge
theory is practically the same thing as the product of a $U(1)$ gauge
theory and an $SU(n)$ gauge theory.  (There is a subtle global difference,
which will be
important later but not in the preliminary remarks that we are
about to make.)  In our problem, the $U(1)$ factor of the gauge group
describes the motion of the string center of mass.  In fact, the $U(1)$
theory has the right degrees of freedom to do that, since in section
two we saw that the $p$-brane world-volume theory at low energies was
simply $U(1)$ supersymmetric gauge theory.  And it is clear that adding
a constant to the position of all $p$-branes shifts the $U(1)$ variable
$\Tr X^i$ without changing the traceless $SU(n)$  part.

In our problem of finding string bound states, since the  $U(1)$ theory
already has massless excitations in correspondence with those of the
elementary string, a supersymmetric $SU(n)$ vacuum that could give
an $SL(2,\Z)$ transform of the elementary string must
have a {\it mass gap},
as any extra massless excitations would spoil the correspondence with
the elementary string.  The possible existence of vacua with a mass gap
is rather delicate and surprising from several points of view.  For
instance, note
that we are here dealing with $N=8$ supersymmetric Yang-Mills
theory, obtained by dimensional reduction from ten dimensions.  In the
$N=2$ or $N=4$ theories, one could immediately exclude the possibility
of a vacuum
with a mass gap because those theories have chiral $R$-symmetries
with anomalous
two point functions which (according to a well-known argument
by 't Hooft) imply
that there can be no mass gap in any vacuum.  For $N=8$,
there is an $SO(8)$ $R$-symmetry that acts chirally in the sense that
the left-moving fermions transform in one spinor representation and the
right-movers in
the other; as these representations have the same quadratic
Casimir, there is no anomaly in the two point function and no immediate
way to disprove on such grounds the existence of a vacuum with a mass gap.

\bigskip\noindent
{\it Topological Sectors}

Since we want ultimately to study bound states of $m$ elementary
strings and $n$ $D$-strings, we must determine precisely what question
in the gauge theory corresponds to such a collection of strings.
It is convenient to answer this by first asking what topological
sectors the $U(n)$ theory has, and then interpreting these for our
problem.

\def\R{{\cal R}}
In general, in two-dimensional gauge theory with any gauge group $G$,
topological sectors can be introduced by placing a charge at infinity
in any desired representation ${\cal R}$.  Because of screening, only
the transformation of ${\cal R} $ under the center of the group really
matters.  For instance, for the free theory, $G=U(1)$ without matter,
we can place at
infinity a particle of any desired  charge $m$.
We then get a state, encountered in section 2, in which the canonical
momentum of the gauge field is $\pi = m$.
For $SU(n)$, with all fields being in the adjoint representation, we can
place at infinity a charge of any desired ``$n$-ality,'' so there are $n$
sectors to consider, as discussed in \ref\oldwitten{E. Witten, ``Theta
Vacua In Two-Dimensional Quantum Chromodynamics,'' Nuovo Cim.
{\bf 51A} (1979) 325.}.

Let us consider our problem with gauge group $U(n)$ (we take at face value
the Chan-Paton structure that indicates that the global structure of the
gauge group is precisely $U(n)$).  Consider what happens if one places
at infinity a ``quark'' in the fundamental representation of $U(n)$.
To see the physical interpretation, consider a vacuum consisting of $n$
widely separated parallel $D$-strings with commuting matrices $X^j$.
This breaks $U(n)$ to $U(1)^n$, one $U(1)$ factor for each $D$-strings.
Any given state of the quark has charge one for one $U(1)$ and zero for
the others, so one of the $n$ $D$-strings is bound with one elementary
string and the others with none.  This topological sector therefore has
the quantum
numbers of $n$ $D$-strings and one elementary one.  By a similar
reasoning, if we put at infinity a tensor product of $m$ quarks, we
get $m$ elementary strings together with $n$ $D$-strings.

As a special case, an antisymmetric combination of $n$ copies of the
fundamental representation is trivial as an $SU(n)$ representation
-- and so does not affect the center of mass or $SU(n)$ part of the theory
at all -- but by shifting the $U(1)$ field strength adds $n$ elementary
quarks.  Thus the center of mass dynamics of $m$ elementary strings
and $n$ $D$-strings depends only on the value of $m$ modulo $n$.
This is one of the predictions of $SL(2,{\bf Z})$, as the matrix
\eqn\hsin{\left(\matrix{ 1 & t \cr 0 & 1\cr }\right),}
with arbitrary integer $t$, can in acting on
\eqn\hugin{\left(\matrix{ m \cr n\cr}\right)}
shift $m$ by any desired multiple of $n$.  This result is analogous
to the fact that in Sen's study of bound states in $N=4$ supersymmetric
Yang-Mills theory in four dimensions \sen, one sees without any detailed
calculation that the
number of BPS-saturated bound states of $m$ electrons and $n$ monopoles
depends only on the value of $m$ modulo $n$.

\bigskip
\noindent
{\it Known Vacua}

Now let us discuss the known supersymmetric vacua of this theory.
The only known vacua, roughly speaking, are the ones in which the
matrices $X^j$ commute, with large and distinct eigenvalues,
breaking $SU(n)$ to $U(1)^{n-1}$.
Supersymmetric non-renormalization theorems imply that the energy is
exactly zero in that region, not just in the classical approximation.

Speaking of a family of vacua is actually  a rough way
of describing the situation;
because of two-dimensional infrared divergences,
the space of eigenvalues of the $X^j$ up to permutation is more like
the target space of a string theory than  the parameter space of
a family of vacua (as it would be if we were discussing $p$-branes of
$p>1$).  At any rate, it is true that -- at least if we do not
introduce a charge at infinity -- the $X^j$ can become large at
no cost in energy, and therefore that the thermodynamic quantities
diverge as if there were a family of vacua.

What happens if there is a charge at infinity?  Since a few tricky
issues will arise, let us begin by considering some special cases.
Suppose that  $G=SU(2)$ and that the representation at infinity
is a quark doublet.  If the $X^j$ go to infinity, $SU(2)$ is broken
to $U(1)$.  The $U(1)$ theory is free at energies low compared
to the mass scale determined by $\langle X^j\rangle $, so subsequent
calculations can be made semiclassically.  Higgsing of $SU(2)$ to $U(1)$
does not lead to screening of the charge at infinity, since every
component of the $SU(2)$ doublet is charged under $U(1)$.  Thus
the $SU(2)$ theory with a quark (or any representation odd under the
center of $SU(2)$) at infinity has the property that there is an
energetic barrier to going to large $X^j$ (or more exactly the energy
at infinity is bounded stricly above the supersymmetric value).
This does not answer
the question of whether the $SU(2)$ theory with a quark at infinity
has a supersymmetric ground state; it merely says that such a state,
if it exists, decays exponentially for large $X^j$.

For an example that gives a different answer, take $n=4$, that
is $G=SU(4)$, and let $ X^j$ go to infinity in a direction
breaking $SU(4)$ to $SU(2)\times SU(2)\times U(1)$.  For $m=1$,
we can take the representation at infinity to be the fundamental
representation ${\cal S}$ of $SU(4)$.  Every component of ${\cal S}$
is charged under the $U(1)$, so for $m=1,\,n=4$, there is an
energetic barrier to going to infinity in this direction (or any
other, as one can easily verify).  Now try $m=2, \,n=4$,
We can take the representation at infinity to be the antisymmetric
product $\wedge^2{\cal S}$
of two copies of ${\cal S}$.  Screening now occurs
as $\wedge^2{\cal S}$ contains a component that is neutral under
$U(1)$.  This component
 transforms as $(2,2)$ under $SU(2)\times SU(2)$, so that
each $SU(2)$ has $m=1$ -- a quark at infinity.  Thus,
we cannot determine without understanding the $SU(2)$  dynamics
-- which is given by a strongly coupled quantum field theory
near the origin -- whether in the $SU(4)$ theory with $m=2$ there
is an energetic barrier to going to large $X^j$.  If -- as we argue
later -- there is a supersymmetric ground state of the $m=1$, $n=2$
system, then there is no energetic barrier to taking $X^j$ to
infinity in this particular direction.

In each case, what we have
described in the gauge theory language is perfectly obvious
in terms of the original $D$-strings.
The discussion of $m=1, n=2$ amounted to saying that (as is obvious
from the BPS formula) if the  $(m,n)= (1,2)$ system is separated
into $(0,1)$ and $(1,1)$ subsystems, then the energy is greater
than the supersymmetric or BPS value for the $(1,2)$ system.
The discussion of the $(2,4) $ case amounted to showing that
 if there is an $m=1, n=2$ supersymmetric
bound state, then there is no energetic barrier to separating
the $(2,4)$ system into two $(1,2)$ subsystems.

Now we discuss the general case.  First we show in gauge theory
language that if $m$ is
prime to $n$, there is an energetic barrier to taking $X^j$ to infinity
in any direction.  Taking $X^j$ to infinity in any direction
will break  $SU(n)$ to a subgroup that contains at least one $U(1)$
factor that can be treated semiclassically; it is enough to show that
one can pick a $U(1)$ for which every component of the representation at
infinity is charged.  Indeed, one can always pick an unbroken $U(1)$ that
has    only two eigenvalues in the fundamental representation ${\cal S}$
of $SU(n)$, say of multiplicity $a$ and $n-a$, with $1\leq a\leq n$.
The eigenvalues are $(n-a)$ and $-a$.  For given $m$, if we place
at infinity the tensor product of $m$ copies of ${\cal S}$,
the possible $U(1)$ eigenvalues are $s(n-a)+(m-s)(-a)$ with
$0\leq s\leq m$; this is congruent to $-ma$ modulo $n$, and so can
never vanish if $m$ is prime to $n$.  Thus the expected energetic
barrier exists when $m$ and $n$ are relatively prime.

For the converse, we must show that if $m$ and $n$ are not
relatively prime, then without knowledge
of strongly coupled non-abelian gauge dynamics, one cannot
prove that there is an energetic barrier to going to large $X$.
Place at infinity the $m^{th}$ antisymmetric tensor
 power of ${\cal S}$.  Let $d$ be the greatest common divisor of
$m$ and $n$.  Take the $X^j$ to infinity in a direction that breaks
$SU(n)$ to $SU(n/d)^d$ times a product of $U(1)$'s.  The
$m^{th}$ antisymmetric power of ${\cal S} $ contains a component
neutral under all of the $U(1)$'s, so without knowledge of
strongly coupled nonabelian gauge dynamics one cannot show
that there is an energetic barrier to taking the $X^j$ to infinity
in the chosen direction (which corresponds to separating the
$(m,n)$ system into $d$ separate $(m/d,n/d)$ subsystems).

\bigskip\noindent
{\it The Perturbation Argument}

\nref\newharv{J. A. Harvey and J. P. Gauntlett, ``$S$-Duality
And The Dyon Spectrum In $N=2$ Super-Yang-Mills Theory,''
hep-th/9508156.}
\nref\sethi{S. Sethi, M. Stern, and E. Zaslow,
``Monopole and Dyon Bound States In $N=2$ Supersymmetric
Yang-Mills Theories,'' hep-th/9508117.}

In what follows, we will use an argument that only works when
there is an energetic barrier to taking the $X^j$ large,
and so only works when $m$ and $n$ are relatively prime.
When $m$ and $n$ are not relatively prime, one would be dealing
-- because the energetic barrier to separation
is absent -- with bound states at threshold, which are extremely
delicate things (though they are possible in quantum mechanics
and duly appear when required by duality, as in \refs{\newharv,\sethi}).
In the present case, as there are apparently
no bound states at threshold of elementary strings,
$SL(2,{\bf Z})$ requires
the absence of bound states whenever $m,n$ are not relatively prime.
But here I will focus only on the relatively prime case in which
the energetic barrier makes things easier.

We have noted  above that the supersymmetric vacua we want for the
relatively prime case should have a mass gap, to avoid   giving
massless world-sheet modes that do not have counterparts for
the elementary string.  It is in fact highly plausible that
any supersymmetric ground state of the $(m,n)$ system in the
relatively prime case would have a mass gap.
We at least know that any such state would have  a true normalizable
vacuum, since there is an energetic barrier  against the $X^j$ becoming
large.  There are no known $N>4$ supersymmetric systems with $SO(N)$
symmetry, massless
particles and a normalizable vacuum; the lack of an $N>4$ superconformal
algebra with unitary representation theory tends to be an obstruction
since it means that there is no conformal field theory for such a
 system to flow to in the infrared.
So we assume that supersymmetric vacua of the $(m,n)$ system all
have a mass gap.  There can be only finitely many of them as none
can be at large $X^i$; the problem is to determine the number.

To argue that the number is one, we use the fact that, once it is
given that there is a mass gap, the number of ground states cannot
change under a small  perturbation. (With massless particles,
it would be possible for a vacuum to split into several under
a weak perturbation, or to break supersymmetry and disappear from the list
of supersymmetric vacua.)
We will simply pick a judicious perturbation that makes things
simple.

\nref\vafa{C. Vafa and E. Witten, ``A Strong Coupling Test Of
$S$-Duality,'' Nucl. PHys. {\bf B431} (1994) 3, hep-th/9408074.}

\nref\donagi{R. Donagi and E. Witten, ``Supersymmetric Yang-Mills
Theory and Integrable Systems,'' hep-th/9510101,      section 4.}
$N=1$ supersymmetric Yang-Mills theory reduces to $N=4$ in four
dimensions.  The $N=4$ theory, viewed as an $N=1$ theory,
has in addition to the vector multiplet three chiral superfields
$A,B,$ and $C$ in the adjoint representation.  The
superpotential is $\Tr\, A[B,C]$.  We make a perturbation to a system
with superpotential
\eqn\lopo{W=\Tr\left(A[B,C]-{\epsilon\over 2}(A^2+B^2+C^2)\right).}
(The special features of this
perturbation were exploited previously in \refs{\vafa,\donagi}.)

Of course, we are really interested in the dimensional reduction
of this system to  two dimensions.  Notice
that, by rescaling of $A,B,C$, and the world-sheet coordinates, and
an $R$-transformation,
operations that only change the Lagrangian by an irrelevant operator
$\int d^2\sigma d^4\theta\dots$, one can adjust $\epsilon$ to any
desired value, given that it is non-zero.  Thus while the mass gap
permits us to introduce a small $\epsilon$ without disturbing
the vacuum structure, we may in fact take $\epsilon$ large
and use semi-classical reasoning.

 The condition for a vacuum state
\eqn\mopo{\eqalign{[A,B]& = \epsilon C\cr
                   [B,C]& = \epsilon A\cr
                   [C,A]& = \epsilon B\cr}}
asserts that the quantities $A/\epsilon $, $B/\epsilon $, and
$C/\epsilon $ obey the commutation relations of $SU(2)$.
At the classical level, a vacuum is given by specifying an $n$-dimensional
representation of $SU(2)$, together with the values of the scalar
fields $\phi,\bar\phi$
that  are in the vector multiplet in four dimensions, but become
scalars in two dimensions.  The two-dimensional scalar potential
has extra terms
\eqn\popo{\Delta V=\sum_{X=A,B,C}\Tr\left( [X,\phi][\bar X,\bar \phi]
+[X,\bar\phi][\bar X,\phi]\right) + \Tr [\phi,\bar \phi]^2,}
 which imply
that  the energy can vanish classically only if
$\phi,\bar\phi$ commute with
each other and with $A,B,C$.

We want   to know whether with a representation at infinity of
$n$-ality  equal to $m$, this system has a supersymmetric ground
state with mass gap at the quantum level.
We can split the discussion into three parts, depending on whether
the $SU(2)$ representation is (i) trivial, (ii) non-trivial but
reducible, or (iii) irreducible.
In the trivial case, as $A,B,$ and $C$ are massive in expanding
around $A=B=C=0$, the low energy theory
is the two-dimensional $N=2$ $SU(n)$ theory (associated with
dimensional reduction from four-dimensional $N=1$).  By an argument
given above, that theory does not have a vacuum with a mass gap\foot{
There is a subtlety here: the argument for absence of mass gap in the
$N=2$ theory depends on a chiral symmetry which is here explicitly
broken by coupling to the massive fields.  The breaking is, however,
irrelevant at low energies unless a twisted chiral superpotential
is generated.  This can be excluded by a variety of standard arguments.
   For instance, the twisted chiral superpotential
would have to be  independent of the mass of the massive
particles, as $\epsilon$ is chiral
rather than     twisted chiral. In  two-dimensional
superrenormalizable theories like
this one, the effects of massive particles on the massless   ones vanish
for large mass, except sometimes for ill-convergent 	one-loop
tadpoles (absent
here because as all fields are in the adjoint representation of $SU(n)$,
the tadpoles  vanish).
So the fact that the twisted chiral superpotential must be independent
of $\epsilon$  means that it cannot be generated at all. }
(and presumably breaks supersymmetry because of the inability to
screen the charge at infinity).
In case (ii), the low energy theory has at least one unbroken
$U(1)$ under which (by an argument given above) no component of the
charge at infinity is neutral, so supersymmetry is spontaneously
broken at the quantum level.

The interesting case therefore is case (iii).  In this case,
the gauge group is completely broken -- screening the charge
at infinity -- and all fields, including
$\phi$ and $\bar \phi$, get a mass.  This therefore is the desired unique
vacuum with mass gap corresponding to the $(m,n)$ string.

\newsec{Bound States Of Other $D$-Branes And $p$-Branes}

In this section, we briefly apply some of the same methods
to other questions involving bound states of $p$-branes.  For
any $p$, one has to look at the dimensional reduction of
ten-dimensional supersymmetric Yang-Mills theory to $p+1$
dimensions.
We first consider the Type IIB theory, and then more superficially
Type IIA.

\subsec{Type IIB}

\bigskip\noindent
{\it Instantons}

One might begin
with the Type IIB
 $-1$-branes or instantons.  They are given by the dimensional
reduction of supersymmetric Yang-Mills theory all the way to
zero dimensions.  Thus, in the $n$-instanton sector, all
ten space-time coordinates become  matrices $X^i,\,i=0,\dots ,9$,
while the fermion zero modes in the instanton sector will be
represented by the supersymmetric partners $\psi_\alpha,\, $
$\alpha=1,\dots,16$, also in the adjoint representation.
The instanton measure is an  integral over $X$ and $\psi$ weighted
by $e^{-I}$, with $I$ a multiple of the dimensionally-reduced
supersymmetric Yang-Mills action:
\eqn\ksnk{I=\sum_{i<j}\Tr [X^i,X^j]^2 +\sum_{i,\alpha,\beta}
\Gamma^{i}_{\alpha\beta}\Tr\psi^\alpha[X^i,\psi^\beta].}
(The $\Gamma$'s are gamma-matrices.)
This is the pure case of seeing the space-time coordinates
as non-commuting matrices.  For a supersymmetric minimum of the
action, of course, we minimize the action with $[X^i,X^j]=0$, and
then the matrices commute.  One can ask whether non-supersymmetric
instantons can be found as higher critical points of $I$ with
non-commuting $X$'s.  A scaling argument shows that there are
none: $I$ scales as $t^4$ under $X\to tX$, so any critical point
has $I=0$.

This description of the effective action for
$-1$-branes has an intriguing similarity to the ADHM description
of Yang-Mills instantons in four dimensions, which are determined
by matrices $X^i$, $ i=1,\dots,4$, obeying not $[X^i,X^j]=0$
but the self-dual counterpart
$[X^i,X^j]={1\over 2}\epsilon^{ijkl}[X_k,X_l]$.

\bigskip\noindent
{\it Three-branes}

We next move on to consider bound states of Dirichlet three-branes.
A bound state with the same world-volume structure as the
basic three-brane would correspond to a
 vacuum  with mass gap of $N=4$ $SU(n)$
supersymmetric Yang-Mills theory in four dimensions.  There are none,
since anomalous triangle diagrams of the
 $SU(4)$ $R$-symmetry currents imply -- by 't Hooft's old argument --
that this theory can have no vacuum with mass gap.
In fact, it is strongly believed that in this theory the moduli
space of vacua is exactly given by the classical
answer.   In three-brane language, this  means that
 the vacua are labeled by the positions
of the $n$ three-branes, up to permutation, so that there are no bound
states even with exotic world-volume structure.

One might think that this result on absence of three-brane bound states
is what is needed to justify the assumption in \strominger\ of
considering            in
conifold physics only simple, and not multiple,
wrapping of three-branes around collapsing three-cycles.  But this would
be too hasty a conclusion; we have argued here only the
absence of three-brane bound-states in flat space, and the conclusion
does not immediately carry over
to the Calabi-Yau context where the three-branes
are wrapped around a curved cycle.  The opposite
result in flat space would, however, have been undesireable; a bound state
in flat space would have led to a bound state in the conifold problem
when the radius of curvature is large enough (larger than the largest
length scale important in the structure of the flat space bound state),
and therefore, by
BPS-saturation, also when it is small.  By studying the soft
modes in $N=4$ super Yang-Mills theory on ${\bf R}\times {\bf S}^3$,
it can be seen that the bound state problem for $n $ three-branes
wrapped around an ${\bf S}^3$ is equivalent to the       question
-- which will not be addressed here --
of whether there are bound states at threshold in the dimensional
reduction to $0+1$ dimensions
of four-dimensional $N=1$ super Yang-Mills theory
with gauge group $SU(n)$.

\bigskip\noindent
{\it Five-Branes}

For Dirichlet five-branes the story is formally the same.
For bound states with the same structure as the elementary $D$-brane,
one must
consider vacua with mass gap
 in six-dimensional supersymmetric Yang-Mills theory.
There can be none because of anomalous four-point functions of the
$SU(2)\times SU(2)$ $R$-symmetry.  One might worry about the
unrenormalizability of six-dimensional super  Yang-Mills theory,
but this issue seems inessential as the string theory provides
some sort of cutoff, and the anomaly argument is an infrared argument.
The unrenormalizability -- and weak coupling in the infrared --
strongly suggest that the moduli space of vacua is given by the
classical answer, and hence that there are also no bound states
with exotic world-volume structure.

\nref\strom{A. Strominger, ``Heterotic Solitons,''
Nucl. Phys. {\bf B343} (1990) 167.}
\nref\cal{C. G. Callan, Jr., J. A. Harvey,
and A. Strominger, ``World-Sheet Approach To Heterotic
Instantons And Solitons,'' Nucl. Phys. {\bf B359} (1991) 611.}
\nref\dlk{M. J. Duff, R. R. Khuri, and J. X. Lu,
``String Solitons,'' Phys. Rept. {\bf 259} (1995) 213-326,1995,
hep-th/9412184. }
But the absence of     bound states of Dirichlet five-branes
with each other is not the whole story for five-branes.  One must
also consider the solitonic five-brane \refs{\strom - \dlk}.
In fact the two sorts of five-brane form an $SL(2,{\bf Z})$ doublet
just like the two kinds of string.  Electric-magnetic duality in ten
dimensions pairs the five-branes with strings.   $SL(2,{\bf Z})$ predicts
the existence of a bound state of $m$ Dirichlet five-branes with
$n$ solitonic ones for every relatively prime pair $m,n$.  The
prediction cannot be fully tested at the moment because the solitonic
five-branes are not sufficiently well understood, but a few
simple remarks are possible.

Note that in string units, the elementary string and $D$-string
have tensions of order $1$ and $1/\lambda$, respectively;
in Einstein units (where duality acts naturally) these become
$\sqrt\lambda$ and $1/\sqrt \lambda$. The
Dirichlet and solitonic five-brane have tensions of order
$1/\lambda$ and $1/\lambda^2$ , respectively, in string units;
in Einstein units the tensions are of order $\sqrt \lambda$ and
$1/\sqrt\lambda$.  There is thus a parallel between strings
and five-branes with the elementary string mapped to the Dirichlet
five-brane and the $D$-string mapped to the solitonic five-brane.
A BPS-saturated bound state of $m$ Dirichlet   five-branes and
$n$ solitonic ones would have tension in string units
\eqn\hins{\tilde T_{m,n}=
{\tilde T\over \lambda}\sqrt{m^2+{n^2\over \lambda^2}},}
a formula quite analogous to \ikko\ for strings.

The first bound state problem that we considered in section two was
a bound state of one $D$-string and $m$ elementary strings.  The
main qualitative issue was that the tension of the elementary string
completely disappears (in the weak coupling limit) in the presence
of of a $D$-string.  There is a similar issue for a bound state of
one solitonic five-brane with $m$ Dirichlet five-branes.  The $m$
Dirichlet five-branes in vacuum
would have a tension  of order $m/\lambda$,
but the above  formula says  that
for weak coupling this energy practically disappears in the field of
a solitonic five-brane.  Adding the $m$ Dirichlet five-branes to the
solitonic one should increase the ground state energy (in string
units) by an amount
only of order $m^2$, and not $m/\lambda$, for small $\lambda$.

How can this be?  The key is that the soliton five-brane is given
explicitly \refs{\strom - \dlk}
by a four-dimensional solution which contains a region
(an infinite tube that represents a sort of hole in space-time) in which
the dilaton blows up and the effective value of $\lambda$  goes to
infinity.  The energy of a Dirichlet five-brane therefore vanishes as
it falls down the hole.  To make this quantitative, one would have
to understand better the strong-coupling region of the soliton,
but one can at least assert that the main surprising feature of the
soliton solution, which is existence of the strong coupling end,
 is just what is needed  to make the $SL(2,{\bf Z})$
prediction possible.

\bigskip\noindent
{\it Seven-branes}

The Type  IIB theory also  has a Dirichlet seven-brane.  Its bound
states with standard world-volume structure
would correspond to vacua with mass gap in the dimensional
reduction of ten-dimensional super Yang-Mills theory to eight dimensions.
Such vacua do not exist, because the $U(1)$ global symmetry
(which arises in the dimensional reduction from ten to eight dimensions)
has an anomalous five point function, which would be impossible in
a vacuum with mass gap.  The weak infrared coupling of the theory
strongly suggests that bound states with exotic world-volume
structure are also absent.

\subsec{Type IIA Superstrings}

Now we move on to the Type IIA superstring in ten dimensions.
There are Dirichlet $p$-branes for $p$ even, while elementary
and solitonic $p$-branes only exist for odd $p$ (in fact, $p$ equal to
one or five).
So the only bound states to inquire about are the bound states of
$D$-branes with themselves, which correspond to
 vacua with mass gap in ten-dimensional $SU(n)$
super Yang-Mills theory dimensionally reduced to $p+1$ dimensions.

Predictions for small $p$  seem to follow from results
about string dynamics.  For $p=0$, where one is dealing with ordinary
particles (carrying Ramond-Ramond electric charge), precisely one
bound state for each $n$
is  apparently needed to agree with the Kaluza-Klein spectrum of
eleven-dimensional supergravity.
For $p=2$, to make sense of the physics of conifolds \strominger,
one wants no bound state when two-branes are wrapped around a
two-cycle, and therefore (taking the limit as the Calabi-Yau manifold
is scaled up) no bound state in flat space.  For $p>2$ there seem to be
no known predictions.

For $p=0$, one has ordinary quantum mechanics, albeit supersymmetric
quantum mechanics of a rather special sort.  One wants to know whether
there are bound states at threshold, and these might be accessible
to analysis, though the question is beyond the reach of the present paper.

For $p>0$, one is dealing with odd-dimensional quantum field theory,
and  anomaly-based arguments to exclude ground states with mass gap are
not nearly as powerful as they are in even dimesions.  In some cases,
however, some results can be obtained  using discrete anomalies.  For
instance, for $p=2$, the relevant three-dimensional
super Yang-Mills theory
has an $SO(7)$ global symmetry (obtained by dimensional reduction from
ten dimensions).  If one weakly gauges the $SO(7)$, then one can consider
whether the effective action of the theory is even or odd under a
topologically non-trivial $SO(7)$ gauge transformation.  For $n$ even
(so that the dimension of $SU(n)$ is odd), reasoning given  on p. 309
 of \ref\alv{L. ALvarez-Gaum\'e and E. Witten, ``Gravitational
Anomalies,'' Nucl. Phys. {\bf B234} (1983) 269.} shows that
 the effective action is odd, behavior that cannot be reproduced
in a parity-conserving theory with mass gap.  (The global anomaly
can be reproduced in a theory with mass gap
 by adding a parity-violating Chern-Simons interaction.)
So any  vacua with mass gap have
spontaneously broken parity and are paired by the action of parity; the
total number of bound states with standard world-volume structure
is therefore even.

\listrefs
\end

\\
Title: Bound States Of Strings And p-Branes
Author: Edward Witten
Comments: 25 pages, harvmac
\\
The recent discovery of an explicit conformal field theory
description of Type II $p$-branes makes it possible to
investigate the existence of bound states of such objects.
In particular, it is possible with reasonable  precision
to verify the prediction that the Type IIB superstring in ten dimensions
has a family of soliton and bound state
strings permuted by $SL(2,{\bf Z})$.  The space-time coordinates
enter tantalizingly in the formalism as non-commuting matrices.
\\
\end

\\
Title: Bound States Of Strings And $p$-Branes
Author: Edward Witten
Comments: Added references
\\
The recent discovery of an explicit conformal field theory
description of Type II $p$-branes makes it possible to
investigate the existence of bound states of such objects.
In particular, it is possible with reasonable  precision
to verify the prediction that the Type IIB superstring in ten dimensions
has a family of soliton and bound state
strings permuted by $SL(2,{\bf Z})$.  The space-time coordinates
enter tantalizingly in the formalism as non-commuting matrices.
\\

\input harvmac
\newcount\figno
\figno=0
\def\fig#1#2#3{
\par\begingroup\parindent=0pt\leftskip=1cm\rightskip=1cm\parindent=0pt
\baselineskip=11pt
\global\advance\figno by 1
\midinsert
\epsfxsize=#3
\centerline{\epsfbox{#2}}
\vskip 12pt
{\bf Fig. \the\figno:} #1\par
\endinsert\endgroup\par
}
\def\figlabel#1{\xdef#1{\the\figno}}
\def\encadremath#1{\vbox{\hrule\hbox{\vrule\kern8pt\vbox{\kern8pt
\hbox{$\displaystyle #1$}\kern8pt}
\kern8pt\vrule}\hrule}}

\overfullrule=0pt

%
\def\tilde{\widetilde}
\def\bar{\overline}
\def\Z{{\bf Z}}

\def\S{{\bf S}}
\def\R{{\bf R}}

\font\zfont = cmss10 

\def\bigone{\hbox{1\kern -.23em {\rm l}}}
\def\ZZ{\hbox{\zfont Z\kern-.4emZ}}

\Title{hep-th/9510135, IASSNS-HEP-95-83}
{\vbox{\centerline{BOUND STATES OF STRINGS AND $p$-BRANES}}}
\smallskip
\centerline{Edward Witten\foot{Research supported in part
by NSF  Grant PHY92-45317.}}
\smallskip
\centerline{\it School of Natural Sciences, Institute for Advanced Study}
\centerline{\it Olden Lane, Princeton, NJ 08540, USA}\bigskip

\medskip

\noindent
The recent discovery of an explicit conformal field theory
description of Type II $p$-branes makes it possible to
investigate the existence of bound states of such objects.
In particular, it is possible with reasonable  precision
to verify the prediction that the Type IIB superstring in ten dimensions
has a family of soliton and bound state
strings permuted by $SL(2,{\bf Z})$.  The space-time coordinates
enter tantalizingly in the formalism as non-commuting matrices.
\Date{October, 1995}
\newsec{Introduction}

\nref\townsend{P. Townsend, ``The Eleven-Dimensional
Super-Membrane Revisited,''   Phys. Lett. {\bf B350} (1995) 184,
hep-th/9501068.}
 \nref\witten{E. Witten,
``String Theory Dynamics In Various Dimensions,''
 Nucl.Phys. {\bf B443} (1995) 85.   hep-th/9503124.}
In many recent developments involving Type II superstrings, particles
and $p$-branes
carrying Ramond-Ramond charges have played an important role.
In many instances it is important to know about possible bound
states of such objects.  For instance, to make sense of the
strong coupling behavior of the Type IIA superstring in ten dimensions
it appears to be necessary to assume that there are Ramond-Ramond or RR
zero-branes of any (quantized) charge \refs{\townsend,\witten}.
On the other hand, to make sense of the behavior of compactified
Type II superstrings near certain conifold singularities it seems
necessary to assume that two-branes and three-branes behave under
certain circumstances as if there are no bound states \ref\strominger{A.
Strominger, ``Massless Black Holes And Conifolds In String
Theory,'' Nucl. Phys. {\bf B451} (1995) 96, hep-th/9504090.}.

\nref\hull{C. Hull and P. Townsend, ``Unity Of Superstring Dualities,''
Nucl. Phys. {\bf B438} (1995) 109, hep-th/9410167.}
One of the most interesting problems of this kind -- and the
main focus in this paper though we will also discuss other
cases -- concerns one-branes or strings in ten dimensions.
The   ten-dimensional Type IIB theory is believed to have an
$SL(2,{\bf Z})$ $S$-duality symmetry \refs{\hull,\witten}.
The two two-forms  of the theory, one from the NS-NS sector and
one from the RR sector, transform as a doublet under
$SL(2,{\bf Z})$.\foot{In general, the massless $p$-forms
of low energy supergravity theories can be assigned to     representations
of the non-compact symmetry groups even though those groups
are spontaneously broken.  Otherwise, parallel transport in the moduli
space of vacua would clash with Dirac quantization of the charges
of $p-1$-branes.}   The elementary Type IIB superstring
   is a source for
the usual $B$-field from the NS-NS sector, and not for  the RR field.
Let us describe this by saying that it has charges $(1,0)$.
$SL(2,{\bf Z})$ will map the elementary string to a string with
charges $(m,n)$ for any relatively prime pair of integers $m,n$.
The $SL(2,{\bf Z})$ prediction\foot{
This prediction is
perhaps somewhat analogous to the $S$-duality
prediction of bound states of electrons and
monopoles in $N=4$ supersymmetric
Yang-Mills theory, as checked in the two monopole sector
by Sen \ref\sen{A. Sen, ``Dyon-Monopole Bound States, Self-dual
Harmonic Forms On The Multi-Monopole Moduli Space, and $SL(2,{\bf Z})$
Invariance In String Theory,'' Phys. Lett. {\bf B329} (1994) 217,
hep-th/9402032.}}
 for the mass of these solitonic
and composite strings has been worked out in detail \ref\schwarz{J.
H. Schwarz, ``An $SL(2,{\bf Z})$ Multiplet Of Type II
Superstrings,'' hep-th/9508143,
``Superstring Dualities,'' hep-th/9509148.}.
The mass formula is  determined by the fact that the extended
strings are BPS-saturated, invariant under half of the supersymmetries
(which is also the reason that one can make sense of how the spontaneously
broken $SL(2,{\bf Z})$ acts on these strings), so it will be enough
in this paper to find a BPS-saturated string for every relatively
prime pair $m,n$ without directly computing the masses.

What makes these questions accessible for study is that an
explicit and amazingly simple description of strings (and more
general $p$-branes) carrying RR charge has recently been given
\ref\polchinski{J. Polchinski,
``Dirichlet Branes And Ramond-Ramond Charges,''  hep-th/9510017.}
in the form of $D$-strings
and $D$-branes.  The basic $D$-string has charges $(0,1)$, since
it was seen in \polchinski\ to carry  RR charge, but on symmetry
grounds
does not carry the NS-NS charge.  In section 2, we show
that the $(p,1)$ bound states exist as an almost immediate consequence
of the $D$-string structure.  To see the $(m,n)$ strings with
$n>1$ requires a more elaborate construction to which we then turn in
section 3.
In the process, two-dimensional gauge theory makes a perhaps
unexpected appearance; the existence of the BPS-saturated $(m,n)$
strings is equivalent to the existence of certain previously
unknown vacua with mass gap in two-dimensional
$N=8$  supersymmetric gauge theory!
A reasonably compelling argument for the existence of these vacua will
be presented.
We then go on in section 4 to discuss bound states
of  some of the other
 $p$-branes.

One of the most interesting       aspects of the formalism is that
the space-time coordinates normal to the $p$-brane world-sheet
-- or all of them for instantons, that is for $p=-1$ -- enter
as non-commuting matrices (in the adjoint representation of $SU(n)$
in the case of a bound state of $n$ $D$-branes).  Though
perhaps easily
motivated  given the relation of $D$-branes to Chan-Paton factors via
$T$-duality
\nref\oldpolch{J. Dai, R. G. Leigh, and J. Polchinski,
``New Connections Between String Theories,'' Mod. Phys. Lett. {\bf A4}
(1989) 2073.}
\nref\otherpolch{J. Polchinski, ``Combinatorics of Boundaries In
String Theory,'' Phys. Rev. {\bf D50} (1994) 6041, hep-th/9407031.}
\nref\horava{P. Horava, ``Strings On World-Sheet Orbifolds,''
Nucl. Phys. {\bf B327} (1989) 461, ``Background Duality Of Open
String Models,'' Phys. Lett. {\bf B321} (1989) 251.}
\nref\green{M. B. Green, ``Space-Time Duality And Dirichlet
String Theory,'' Phys. Lett. {\bf B266} (1991) 325.}
\nref\sagnotti{A. Sagnotti, ``Open Strings And Their Symmetry
Groups,'' in Cargese `87, ``Nonperturbative Quantum Field Theory,''
ed. G. Mack et. al. (Pergamon Press, 1988) p. 527.}
\refs{\oldpolch - \sagnotti}, this seems possibly
significant for the general understanding of string theory.

\newsec{The $(n,1)$ Strings}

First we recall the structure of $D$-branes, and verify that the
$D$-string has the same low-lying world-sheet excitations as the
fundamental Type IIB superstring and so can be interpreted as
the desired $(0,1)$ string.

We work in ten-dimensional Minkowski space, with a time coordinate
$x^0$ and space coordinates $x^1,\dots ,x^9$.  Ordinarily the
Type II theory has closed strings only.  The Dirichlet $p$-brane
-- with $p$ even or odd for Type IIA or Type IIB --
is simply an object whose presence modifies the allowed boundary
conditions of strings so that in addition  to the usual Neumann
boundary conditions, one may also have Dirichlet boundary conditions
with (for example) $x^{p+1}=\dots=x^9=0$.  How to compute the mass per
unit volume
of a $p$-brane was explained in \otherpolch, while the fact that
the $p$-brane couples to the appropriate RR $p+1$-form is
explained in \polchinski.

Once such  an object is introduced, in  addition to the usual
closed strings, one can have open strings whose ends are at any
values of $x^0,\dots,x^{p}$ with $x^j=0$ for $j>p$.
These strings describe the excitations of the $p$-brane.  The quantization
of these open strings is isomorphic to the conventional quantization
of an oriented open superstring.  The massless states are a vector and
a spinor making up a ten-dimensional supersymmetric Yang-Mills
multiplet with gauge group $U(1)$.
As the zero modes of $ x^j,\,j>p$ are eliminated
by the boundary conditions, the massless particles are functions
only of $x^0,\dots,x^p$.  The massless bosons $A_i(x^s)$, $i,s=0,\dots
p$ propagate as a $U(1)$ gauge boson on the $p$-brane world-surface, while
the other components $\phi_j(x^s)$ with  $j>p$, $s=0,\dots,p$,
 are scalars in the $p+1$-dimensional
sense.  Note that the vectors have conventional open-string gauge
boson vertex operators $V_A=\sum_{i=0}^p
A_i(X^s)\partial_\tau X^i$, with $\partial_\tau$ the
derivative   tangent to the world-sheet boundary, while (as in
\otherpolch) the
scalars have vertex operators of the form $V_\phi=\sum_{j>p}\phi_j(X^s)
\partial_\sigma X^j$ with $\partial_\sigma$ the normal derivative
to the boundary.   For $\phi_j=\,{\rm constant}$, the boundary integral
of $V_\phi$ is the change in the world-sheet action upon adding a constant
to $X^j,\,j>p$, so the scalars can be interpreted as oscillations
in the position of the $p+1$-brane.  This interpretation is
developed in some detail in \ref\leigh{R. Leigh,
``Dirac-Born-Infeld Action From Dirichlet Sigma Model,''
Mod. Phys. Lett. {\bf A4} (1989) 2767.}.
The theory on the $p+1$-dimensional world-volume is naturally
thought of as the ten-dimensional $U(1)$ supersymmetric gauge theory
dimensionally reduced to $p+1$ dimensions.

\def\Z{{\bf Z}}
In particular, consider the one-brane or $D$-string of the Type IIB
theory, in the conventional Lorentz background with the
RR scalar vanishing.  In this case the $U(1)$ gauge field $A_i,\,
i=0,1$ describes no propagating degrees of freedom, though it
will nonetheless play an important role latter.  The massless
bosons are the transverse oscillations of the string.  As for
the massless fermions, after the GSO projection they are a spinor
$\psi_\alpha,$ $\alpha = 1\dots 16$, with definite chirality in the
ten-dimensional sense.  The left and right-moving fermions on the
$D$-brane world-sheet are thus components of ten-dimensional
spinors of the {\it same} chirality (they are even components
of the same spinor), as is usual for Type IIB.  Thus,   the
$D$-string has, as expected, the same world-sheet structure
as the elementary Type IIB superstring, making possible its
interpretation as the $(0,1)$ $SL(2,{\bf Z})$ partner of the $(1,0)$
elementary string.

\bigskip \noindent
{\it Bound States Of Fundamental Strings With A $D$-String}

Now we begin our study of bound states by looking for the
$(m,1)$ states, that is, the bound states of $m$ fundamental
strings with a $D$-string.  To see that something special
must be at work, it is enough to consult the expected mass formula
\schwarz.  For the case in which the RR scalar vanishes, the
tension in string units of an $(m,n)$ string is
\eqn\ikko{T_{m,n}=T\sqrt{ m^2+{n^2\over\lambda^2}} }
with $T$ a constant and $\lambda$ the string coupling constant.
In particular, $T_{0,1}\sim1/\lambda$, an example of
the fact that RR charges have mass
of order $1/\lambda $ in string units.

The point about \ikko\ which is perhaps surprising at first
sight is that the binding energy of an elementary string with
a $D$-string to form a $(1,1)$ string is almost one hundred
per cent.  In fact,
an elementary string has a tension of order one,
but the difference in tension between a $(0,1)$ $D$-string
and a $(1,1)$ string is of order $\lambda$, so to first order
the tension energy of an elementary string completely disappears
when it combines with a $D$-string.

\def\R{{\bf R}}
\def\S{{\bf S}}
While this may sound surprising, it is easy to see in the $D$-string
picture why  it is so.
To make the discussion clean, compactify the $x^1$ direction,
so that we are working on $\R\times \S^1\times \R^8$, where
$\R$ is the time direction, $x^1$ is the spatial direction of
the strings we will consider,
and $\R^8$ encompasses the normal directions.
Consider first, in the absence of $D$-strings, a collection
of ordinary strings wrapping around $\S^1$.  The total string winding
number is conserved because the closed strings are not permitted
to break, or alternatively because after integrating
over $\S^1$ there is a gauge field on
the non-compact low energy world $\R\times \R^8$
-- namely the component $B_{i\,1}$ of the Neveu-Schwarz $B$ field --
whose conserved charge is the total winding number of strings.

Now suppose that there is also a $D$-string wrapping around the $\S^1$,
described as above by allowing strings to terminate at
$x^2=\dots = x^9=0$.  In the presence of such a $D$-string,
the winding number of elementary strings around the $\S^1$ is not
conserved!  In fact, let $V_W$ be the vertex operator of an elementary
string wrapping a certain number of times around the $\S^1$.  In
the presence of the $D$-string, $V_W$ has a non-zero one point function
on the disc, because the boundary of the disc, while fixed
at $x^j=0,\,j>1$, is permitted to wrap around $\S^1$ any required
number of times.

\def\F{{\cal F}}
Since there is a conserved gauge charge coupled to the
elementary string winding states, how can it be that the winding
states can disappear in the $D$-brane sector?  Obviously, the
conserved charge must be carried also by some     degrees of freedom
on the $ D$-brane world-sheet.  This in fact happens as a result
of a mechanism that has long been known
\ref\cremmer{E. Cremmer and J. Scherk, ``Spontaneous Dynamical
Breaking Of Gauge Symmetry In Dual Models,''
 Nucl. Phys. {\bf B72} (1974) 117.}
in the context of
oriented open and closed bosonic strings.
The Neveu-Schwarz $B$ field couples in bulk to the string world-sheet
$\Sigma$ by an interaction
\eqn\huggo{I_B={1\over 2}
\int_\Sigma d^2\sigma\epsilon^{\alpha\beta}\partial_\alpha
X^i\partial_\beta X^j B_{ij},}
Under a gauge transformation $B_{ij}\to B_{ij}+\partial_i\Lambda_j
-\partial_j\lambda_i$, \huggo\ is invariant if the boundary of
$\Sigma$ is empty.  If not, then $I_B$ transforms as
\eqn\kuggo{I_B\to I_B+\int_{\partial\Sigma}d\tau \,\,\Lambda_i{dX^i\over
d\tau}.}
Before concluding that the theory with the $B$ field is inconsistent
in the presence of boundaries, one must remember that the oriented
open string has a $U(1)$ gauge field $A$, encountered above,
which couples to the boundary by
\eqn\juggo{ I_A=\int_{\partial\Sigma}d\tau A_i {dX^i\over d\tau}.}
Gauge invariance is therefore restored if the gauge transformation of
$B$ is accompanied by $A\to A-\Lambda$.  The gauge-invariant
field strength is therefore not $F_{ij}=\partial_iA_j-\partial_jA_i$
but $ \F_{ij}=F_{ij}-B_{ij}$.

In the case at hand,
the $B$ kinetic energy is an integral over all space ${\cal M}$,
while
the  gauge field ``lives'' only on the $D$-brane world-sheet $\Sigma$.
The  relevant part of the effective action therefore scales  like
\eqn\pluggo{I=\int_{\cal M}d^{10}x\sqrt g {1\over 2\lambda^2}|dB|^2
              +\int_\Sigma d^2x \sqrt{g_\Sigma}{1\over 2\lambda}
{\cal F}^2}
where now ${\cal F}=\epsilon^{ij}{\cal F}_{ij}/2$ and $g,g_\Sigma$
are the metrics on ${\cal M}$ and on $\Sigma$.  The scaling
with $\lambda$ comes from the fact that the first term comes
from the two-sphere, and the second from the disc.

A gauge field in two dimensions has no propagating degrees of freedom,
and indeed the equation of motion of $A$ asserts that ${\cal F}$ is
constant.  The equation of motion for $B$ contains ${\cal F}$ as a source,
so $D$-brane states with non-zero ${\cal F}$ carry the $B$-charge;
these are the sought-for $D$-brane states carrying the charge that
in the absence of $D$-branes is carried only by string winding states.

The momentum conjugate to $A$ is $\pi = {\cal F}/\lambda$,
and as we will see momentarily $\pi$ is quantized in integer units,
so ${\cal F}=m\lambda$ for integer $m$. With this value of ${\cal F}$,
the ${\cal F}$ dependent part of the energy is of order
 of order $ m^2\lambda$ as expected from \ikko.

Quantization of $\pi$ as claimed above is equivalent to the statement
that the gauge group is $U(1)$ rather than ${\bf R}$.  The Hamiltonian
constraints of      two-dimensional gauge theory require
that the states  be invariant under infinitesimal
gauge transformations and so are linear combinations of the states
$\psi_m(A)=\exp(im\int_{{\bf S}^1} A)$.   The  state $\psi_m$
 -- which
has $\pi=m$ --  is only invariant under topologically non-trivial $U(1)$
gauge transformations if $m$ is an integer.  It must be the case
that the gauge group is $U(1) $ and $m$ is quantized, because if
$D$-strings could carry a continuum of values of the elementary
string winding number, then  annihilation of $D$-strings and
antistrings
would yield an inexistent continuum of string winding number states
in the vacuum sector.\foot{Though the principles for $p$-forms
of $p>1$ are less clear, an analogous mechanism involving compactness
of the gauge group may well cause the vacuum parameter discussed
at the end of \polchinski\ to be quantized.}

If one  wants the allowed values of $\pi$ to be not integers $m$, but
rather to be of the form $m+\theta/2\pi$ with some fixed $\theta$, then
the method to achieve that is well known \ref\coleman{S. Coleman,
``More About The Massive Schwinger Model,'' Ann. Phys. {\bf 101} (1976)
239.}.
One must add to the action a term of the form $\int_\Sigma
{\rm constant}\cdot {\cal F}$.  In our problem the ``constant''
must be the scalar field $\beta$ from the RR sector; the desired coupling
should come from a $\beta - A$ two point function on the disc.
This shift in the allowed values of the elementary string winding
number for $D$-strings appears in the general $SL(2,{\bf Z})$-invariant
formula \schwarz\ for the tension of $D$-strings.  It is analogous
to the $\theta$-dependent shift in the allowed  electric   charges of
a magnetic monopole \ref\oldwitten{E. Witten, ``Dyons Of Charge
$e\theta/2\pi$,''  Phys. Lett. {\bf 86B} (1979) 283.}.

There actually is a good   analogy between the problem we have
just analyzed and certain questions about monopoles in, say,
$N=4$ super Yang-Mills theory in four dimensions. In that theory,
the BPS formula for masses of electrons and dyons shows that
for weak coupling the rest mass of an electron nearly disappears
when it combines with a monopole.  That occurs because the classical
monopole solution is not invariant under charge rotations;
there is therefore a collective coordinate which for weak coupling
can carry charge at very little cost in energy.  For strings,
the gauge transformation that measures string winding number is
a constant mode of the gauge parameter $\Lambda$ of the $B$ field.
The equation $\delta A=-\Lambda$ shows that the $D$-string
is not invariant under this gauge transformation, so that there
is a collective coordinate -- namely $\int_{{\bf S}^1} A $ --
that can carry string winding number at very little cost in energy.

\newsec{Bound States Of $D$-Strings}

We will here describe the general setting for analyzing bound
states of $D$-branes and then apply it to $D$-strings.

First of all, to find a bound state of $n$ parallel Dirichlet $p$-branes,
one needs a description of the relevant low energy physics
that is valid when the branes are nearby.
Because of the relation of $D$-branes to Chan-Paton factors,
there is an obvious guess for what happens when the  $D$-branes
are precisely on top of each other: this should correspond to
an unbroken $U(n)$ gauge symmetry, with the effective action
being ten-dimensional $U(n)$ supersymmetric gauge theory,
dimensionally reduced to the $p+1$-dimensional world-volume of
the $D$-brane.

This ansatz can be justified by considering two parallel Dirichlet
$p$-branes,
say one at $x^j=0$, $j>p$, and one at $x^j=a^j$, $j>p$.  To find
the excitation spectrum in the presence of two $p$-branes, in addition
to closed strings, we must consider strings that start and end on
one or the other $p$-brane.  Strings  that start and end on the
same brane give a $U(1)\times U(1)$ gauge theory (with one $U(1)$
living one each $p$-brane) as we have discussed at the beginning
of section 2.  We must also consider strings starting at the first
brane and ending at the second, or vice versa.  Such strings
have $U(1)\times U(1)$ charges $(-1,1)$ or $(1,-1)$, respectively.
The ground state in this sector (as usual for supersymmetric open
strings) is a vector multiplet; as
 the string must stretch from the origin to $a=(a^{p+1},\dots,
a^9)$, the vector multiplet mass,
which is the ground state energy in this sector, is $T|a|$ with
$T$ the elementary string tension.  Thus as $a\to 0$, world-volume
vector bosons of charges $(\mp 1, \pm  1)$ becomes massless, giving
the restoration of $U(2)$  gauge symmetry on the world-volume.
Likewise, as $n$ parallel
branes become coincident, one has restoration of a $U(n)$ gauge
symmetry on the world-volume of the $p$-brane.  Filling out the
vector multiplets, the low energy theory on the world-volume
is simply the dimensional reduction of ten-dimensional supersymmetric
Yang-Mills theory with gauge group $U(n)$ from ten to $p+1$ dimensions.

To make the physical interpretation somewhat clearer, let us consider
in some detail the bosonic fields of this theory.  In addition
to the world-volume $U(n)$ gauge field $A$, there are $9-p$ scalar
fields in the adjoint representation of $U(n)$.
Changing notation slightly
from section 2, we will call them $X^j$, $j=p+1,\dots ,9$, because
of their interpretation, which will be clear momentarily, as position
coordinates of the $D$-branes.
The potential energy for the $X^j$ in the dimensional reduction
of ten-dimensional supersymmetric Yang-Mills theory is
\eqn\murf{V={T^2\over 2}\sum_{i,j=p+1}^9 \Tr\,[X^i,X^j]^2.}
Classical states  of unbroken supersymmetry
or zero energy thus have $[X^i,X^j]=0$, so the $X^i$ can
be simultaneously diagonalized, giving $ X^i={\rm diag}(a^i_{(1)},
a^i_{(2)},\dots,a^i_{(n)}).$  We interpret such a configuration as
having $n$ parallel  $p$-branes with the position of the
$\lambda^{th}$ $p$-brane, $\lambda=1,\dots, n$ being the space-time
point $a_\lambda$ with coordinates
$a^i_{(\lambda)}$.  As support for this interpretation, note
that the masses obtained by expanding \murf\ around the given
classical solution are just $T|a_\lambda-a_\mu|,$
$1\leq \lambda<\mu\leq n$, corresponding to strings with one
end at $a_\lambda$ and one at $a_\mu$.  Note that the diagonalization
of the $X^j$ is not unique; the Weyl group of $U(n)$ acts by permuting
the $a_\lambda$'s, corresponding to the fact that the $n$ $p$-branes
are identical bosons (or fermions).

In sum, then, the naive configuration space of $n$ parallel, identical
$p$-branes should be interpreted as the moduli space of classical
vacua of the supersymmetric $U(n)$ gauge theory.  When the $p$-branes
are nearby, the massive modes cannot be ignored; one must then
study the full-fledged quantum $U(n)$ gauge theory, and not
just the slow motion on the classical moduli space of vacua.
The appearance of non-commuting matrices $X^j$ which can be interpreted
as space-time coordinates to the extent that they commute is
highly intriguing for the future.  In what follows, we use this
formalism to study string and $p$-brane bound states.

\bigskip\noindent
{\it The Mass Gap}

In fact, a BPS-saturated  state of $n$ $p$-branes would correspond
simply to a supersymmetric ground state of the effective Hamiltonian,
that is to a supersymmetric vacuum of the $p+1$-dimensional supersymmetric
gauge theory.  Let us discuss just what kind of vacuum would correspond
to a {\it bound state}.

A {\it bound state} is a state that is normalizable except for the
center of mass motion.  In the problem at hand, there is no difficulty
in separating out the overall motion of the center
of mass.  A $U(n)$ gauge
theory is practically the same thing as the product of a $U(1)$ gauge
theory and an $SU(n)$ gauge theory.  (There is a subtle global difference,
which will be
important later but not in the preliminary remarks that we are
about to make.)  In our problem, the $U(1)$ factor of the gauge group
describes the motion of the string center of mass.  In fact, the $U(1)$
theory has the right degrees of freedom to do that, since in section
two we saw that the $p$-brane world-volume theory at low energies was
simply $U(1)$ supersymmetric gauge theory.  And it is clear that adding
a constant to the position of all $p$-branes shifts the $U(1)$ variable
$\Tr X^i$ without changing the traceless $SU(n)$  part.

In our problem of finding string bound states, since the  $U(1)$ theory
already has massless excitations in correspondence with those of the
elementary string, a supersymmetric $SU(n)$ vacuum that could give
an $SL(2,\Z)$ transform of the elementary string must
have a {\it mass gap},
as any extra massless excitations would spoil the correspondence with
the elementary string.  The possible existence of vacua with a mass gap
is rather delicate and surprising from several points of view.  For
instance, note
that we are here dealing with $N=8$ supersymmetric Yang-Mills
theory, obtained by dimensional reduction from ten dimensions.  In the
$N=2$ or $N=4$ theories, one could immediately exclude the possibility
of a vacuum
with a mass gap because those theories have chiral $R$-symmetries
with anomalous
two point functions which (according to a well-known argument
by 't Hooft) imply
that there can be no mass gap in any vacuum.  For $N=8$,
there is an $SO(8)$ $R$-symmetry that acts chirally in the sense that
the left-moving fermions transform in one spinor representation and the
right-movers in
the other; as these representations have the same quadratic
Casimir, there is no anomaly in the two point function and no immediate
way to disprove on such grounds the existence of a vacuum with a mass gap.

\bigskip\noindent
{\it Topological Sectors}

Since we want ultimately to study bound states of $m$ elementary
strings and $n$ $D$-strings, we must determine precisely what question
in the gauge theory corresponds to such a collection of strings.
It is convenient to answer this by first asking what topological
sectors the $U(n)$ theory has, and then interpreting these for our
problem.

\def\R{{\cal R}}
In general, in two-dimensional gauge theory with any gauge group $G$,
topological sectors can be introduced by placing a charge at infinity
in any desired representation ${\cal R}$.  Because of screening, only
the transformation of ${\cal R} $ under the center of the group really
matters.  For instance, for the free theory, $G=U(1)$ without matter,
we can place at
infinity a particle of any desired  charge $m$.
We then get a state, encountered in section 2, in which the canonical
momentum of the gauge field is $\pi = m$.
For $SU(n)$, with all fields being in the adjoint representation, we can
place at infinity a charge of any desired ``$n$-ality,'' so there are $n$
sectors to consider, as discussed in \ref\oldwitten{E. Witten, ``Theta
Vacua In Two-Dimensional Quantum Chromodynamics,'' Nuovo Cim.
{\bf 51A} (1979) 325.}.

Let us consider our problem with gauge group $U(n)$ (we take at face value
the Chan-Paton structure that indicates that the global structure of the
gauge group is precisely $U(n)$).  Consider what happens if one places
at infinity a ``quark'' in the fundamental representation of $U(n)$.
To see the physical interpretation, consider a vacuum consisting of $n$
widely separated parallel $D$-strings with commuting matrices $X^j$.
This breaks $U(n)$ to $U(1)^n$, one $U(1)$ factor for each $D$-strings.
Any given state of the quark has charge one for one $U(1)$ and zero for
the others, so one of the $n$ $D$-strings is bound with one elementary
string and the others with none.  This topological sector therefore has
the quantum
numbers of $n$ $D$-strings and one elementary one.  By a similar
reasoning, if we put at infinity a tensor product of $m$ quarks, we
get $m$ elementary strings together with $n$ $D$-strings.

As a special case, an antisymmetric combination of $n$ copies of the
fundamental representation is trivial as an $SU(n)$ representation
-- and so does not affect the center of mass or $SU(n)$ part of the theory
at all -- but by shifting the $U(1)$ field strength adds $n$ elementary
quarks.  Thus the center of mass dynamics of $m$ elementary strings
and $n$ $D$-strings depends only on the value of $m$ modulo $n$.
This is one of the predictions of $SL(2,{\bf Z})$, as the matrix
\eqn\hsin{\left(\matrix{ 1 & t \cr 0 & 1\cr }\right),}
with arbitrary integer $t$, can in acting on
\eqn\hugin{\left(\matrix{ m \cr n\cr}\right)}
shift $m$ by any desired multiple of $n$.  This result is analogous
to the fact that in Sen's study of bound states in $N=4$ supersymmetric
Yang-Mills theory in four dimensions \sen, one sees without any detailed
calculation that the
number of BPS-saturated bound states of $m$ electrons and $n$ monopoles
depends only on the value of $m$ modulo $n$.

\bigskip
\noindent
{\it Known Vacua}

Now let us discuss the known supersymmetric vacua of this theory.
The only known vacua, roughly speaking, are the ones in which the
matrices $X^j$ commute, with large and distinct eigenvalues,
breaking $SU(n)$ to $U(1)^{n-1}$.
Supersymmetric non-renormalization theorems imply that the energy is
exactly zero in that region, not just in the classical approximation.

Speaking of a family of vacua is actually  a rough way
of describing the situation;
because of two-dimensional infrared divergences,
the space of eigenvalues of the $X^j$ up to permutation is more like
the target space of a string theory than  the parameter space of
a family of vacua (as it would be if we were discussing $p$-branes of
$p>1$).  At any rate, it is true that -- at least if we do not
introduce a charge at infinity -- the $X^j$ can become large at
no cost in energy, and therefore that the thermodynamic quantities
diverge as if there were a family of vacua.

What happens if there is a charge at infinity?  Since a few tricky
issues will arise, let us begin by considering some special cases.
Suppose that  $G=SU(2)$ and that the representation at infinity
is a quark doublet.  If the $X^j$ go to infinity, $SU(2)$ is broken
to $U(1)$.  The $U(1)$ theory is free at energies low compared
to the mass scale determined by $\langle X^j\rangle $, so subsequent
calculations can be made semiclassically.  Higgsing of $SU(2)$ to $U(1)$
does not lead to screening of the charge at infinity, since every
component of the $SU(2)$ doublet is charged under $U(1)$.  Thus
the $SU(2)$ theory with a quark (or any representation odd under the
center of $SU(2)$) at infinity has the property that there is an
energetic barrier to going to large $X^j$ (or more exactly the energy
at infinity is bounded stricly above the supersymmetric value).
This does not answer
the question of whether the $SU(2)$ theory with a quark at infinity
has a supersymmetric ground state; it merely says that such a state,
if it exists, decays exponentially for large $X^j$.

For an example that gives a different answer, take $n=4$, that
is $G=SU(4)$, and let $ X^j$ go to infinity in a direction
breaking $SU(4)$ to $SU(2)\times SU(2)\times U(1)$.  For $m=1$,
we can take the representation at infinity to be the fundamental
representation ${\cal S}$ of $SU(4)$.  Every component of ${\cal S}$
is charged under the $U(1)$, so for $m=1,\,n=4$, there is an
energetic barrier to going to infinity in this direction (or any
other, as one can easily verify).  Now try $m=2, \,n=4$,
We can take the representation at infinity to be the antisymmetric
product $\wedge^2{\cal S}$
of two copies of ${\cal S}$.  Screening now occurs
as $\wedge^2{\cal S}$ contains a component that is neutral under
$U(1)$.  This component
 transforms as $(2,2)$ under $SU(2)\times SU(2)$, so that
each $SU(2)$ has $m=1$ -- a quark at infinity.  Thus,
we cannot determine without understanding the $SU(2)$  dynamics
-- which is given by a strongly coupled quantum field theory
near the origin -- whether in the $SU(4)$ theory with $m=2$ there
is an energetic barrier to going to large $X^j$.  If -- as we argue
later -- there is a supersymmetric ground state of the $m=1$, $n=2$
system, then there is no energetic barrier to taking $X^j$ to
infinity in this particular direction.

In each case, what we have
described in the gauge theory language is perfectly obvious
in terms of the original $D$-strings.
The discussion of $m=1, n=2$ amounted to saying that (as is obvious
from the BPS formula) if the  $(m,n)= (1,2)$ system is separated
into $(0,1)$ and $(1,1)$ subsystems, then the energy is greater
than the supersymmetric or BPS value for the $(1,2)$ system.
The discussion of the $(2,4) $ case amounted to showing that
 if there is an $m=1, n=2$ supersymmetric
bound state, then there is no energetic barrier to separating
the $(2,4)$ system into two $(1,2)$ subsystems.

Now we discuss the general case.  First we show in gauge theory
language that if $m$ is
prime to $n$, there is an energetic barrier to taking $X^j$ to infinity
in any direction.  Taking $X^j$ to infinity in any direction
will break  $SU(n)$ to a subgroup that contains at least one $U(1)$
factor that can be treated semiclassically; it is enough to show that
one can pick a $U(1)$ for which every component of the representation at
infinity is charged.  Indeed, one can always pick an unbroken $U(1)$ that
has    only two eigenvalues in the fundamental representation ${\cal S}$
of $SU(n)$, say of multiplicity $a$ and $n-a$, with $1\leq a\leq n$.
The eigenvalues are $(n-a)$ and $-a$.  For given $m$, if we place
at infinity the tensor product of $m$ copies of ${\cal S}$,
the possible $U(1)$ eigenvalues are $s(n-a)+(m-s)(-a)$ with
$0\leq s\leq m$; this is congruent to $-ma$ modulo $n$, and so can
never vanish if $m$ is prime to $n$.  Thus the expected energetic
barrier exists when $m$ and $n$ are relatively prime.

For the converse, we must show that if $m$ and $n$ are not
relatively prime, then without knowledge
of strongly coupled non-abelian gauge dynamics, one cannot
prove that there is an energetic barrier to going to large $X$.
Place at infinity the $m^{th}$ antisymmetric tensor
 power of ${\cal S}$.  Let $d$ be the greatest common divisor of
$m$ and $n$.  Take the $X^j$ to infinity in a direction that breaks
$SU(n)$ to $SU(n/d)^d$ times a product of $U(1)$'s.  The
$m^{th}$ antisymmetric power of ${\cal S} $ contains a component
neutral under all of the $U(1)$'s, so without knowledge of
strongly coupled nonabelian gauge dynamics one cannot show
that there is an energetic barrier to taking the $X^j$ to infinity
in the chosen direction (which corresponds to separating the
$(m,n)$ system into $d$ separate $(m/d,n/d)$ subsystems).

\bigskip\noindent
{\it The Perturbation Argument}

\nref\newharv{J. A. Harvey and J. P. Gauntlett, ``$S$-Duality
And The Dyon Spectrum In $N=2$ Super-Yang-Mills Theory,''
hep-th/9508156.}
\nref\sethi{S. Sethi, M. Stern, and E. Zaslow,
``Monopole and Dyon Bound States In $N=2$ Supersymmetric
Yang-Mills Theories,'' hep-th/9508117.}

In what follows, we will use an argument that only works when
there is an energetic barrier to taking the $X^j$ large,
and so only works when $m$ and $n$ are relatively prime.
When $m$ and $n$ are not relatively prime, one would be dealing
-- because the energetic barrier to separation
is absent -- with bound states at threshold, which are extremely
delicate things (though they are possible in quantum mechanics
and duly appear when required by duality, as in \refs{\newharv,\sethi}).
In the present case, as there are apparently
no bound states at threshold of elementary strings,
$SL(2,{\bf Z})$ requires
the absence of bound states whenever $m,n$ are not relatively prime.
But here I will focus only on the relatively prime case in which
the energetic barrier makes things easier.

We have noted  above that the supersymmetric vacua we want for the
relatively prime case should have a mass gap, to avoid   giving
massless world-sheet modes that do not have counterparts for
the elementary string.  It is in fact highly plausible that
any supersymmetric ground state of the $(m,n)$ system in the
relatively prime case would have a mass gap.
We at least know that any such state would have  a true normalizable
vacuum, since there is an energetic barrier  against the $X^j$ becoming
large.  There are no known $N>4$ supersymmetric systems with $SO(N)$
symmetry, massless
particles and a normalizable vacuum; the lack of an $N>4$ superconformal
algebra with unitary representation theory tends to be an obstruction
since it means that there is no conformal field theory for such a
 system to flow to in the infrared.
So we assume that supersymmetric vacua of the $(m,n)$ system all
have a mass gap.  There can be only finitely many of them as none
can be at large $X^i$; the problem is to determine the number.

To argue that the number is one, we use the fact that, once it is
given that there is a mass gap, the number of ground states cannot
change under a small  perturbation. (With massless particles,
it would be possible for a vacuum to split into several under
a weak perturbation, or to break supersymmetry and disappear from the list
of supersymmetric vacua.)
We will simply pick a judicious perturbation that makes things
simple.

\nref\vafa{C. Vafa and E. Witten, ``A Strong Coupling Test Of
$S$-Duality,'' Nucl. PHys. {\bf B431} (1994) 3, hep-th/9408074.}

\nref\donagi{R. Donagi and E. Witten, ``Supersymmetric Yang-Mills
Theory and Integrable Systems,'' hep-th/9510101,      section 4.}
$N=1$ supersymmetric Yang-Mills theory reduces to $N=4$ in four
dimensions.  The $N=4$ theory, viewed as an $N=1$ theory,
has in addition to the vector multiplet three chiral superfields
$A,B,$ and $C$ in the adjoint representation.  The
superpotential is $\Tr\, A[B,C]$.  We make a perturbation to a system
with superpotential
\eqn\lopo{W=\Tr\left(A[B,C]-{\epsilon\over 2}(A^2+B^2+C^2)\right).}
(The special features of this
perturbation were exploited previously in \refs{\vafa,\donagi}.)

Of course, we are really interested in the dimensional reduction
of this system to  two dimensions.  Notice
that, by rescaling of $A,B,C$, and the world-sheet coordinates, and
an $R$-transformation,
operations that only change the Lagrangian by an irrelevant operator
$\int d^2\sigma d^4\theta\dots$, one can adjust $\epsilon$ to any
desired value, given that it is non-zero.  Thus while the mass gap
permits us to introduce a small $\epsilon$ without disturbing
the vacuum structure, we may in fact take $\epsilon$ large
and use semi-classical reasoning.

 The condition for a vacuum state
\eqn\mopo{\eqalign{[A,B]& = \epsilon C\cr
                   [B,C]& = \epsilon A\cr
                   [C,A]& = \epsilon B\cr}}
asserts that the quantities $A/\epsilon $, $B/\epsilon $, and
$C/\epsilon $ obey the commutation relations of $SU(2)$.
At the classical level, a vacuum is given by specifying an $n$-dimensional
representation of $SU(2)$, together with the values of the scalar
fields $\phi,\bar\phi$
that  are in the vector multiplet in four dimensions, but become
scalars in two dimensions.  The two-dimensional scalar potential
has extra terms
\eqn\popo{\Delta V=\sum_{X=A,B,C}\Tr\left( [X,\phi][\bar X,\bar \phi]
+[X,\bar\phi][\bar X,\phi]\right) + \Tr [\phi,\bar \phi]^2,}
 which imply
that  the energy can vanish classically only if
$\phi,\bar\phi$ commute with
each other and with $A,B,C$.

We want   to know whether with a representation at infinity of
$n$-ality  equal to $m$, this system has a supersymmetric ground
state with mass gap at the quantum level.
We can split the discussion into three parts, depending on whether
the $SU(2)$ representation is (i) trivial, (ii) non-trivial but
reducible, or (iii) irreducible.
In the trivial case, as $A,B,$ and $C$ are massive in expanding
around $A=B=C=0$, the low energy theory
is the two-dimensional $N=2$ $SU(n)$ theory (associated with
dimensional reduction from four-dimensional $N=1$).  By an argument
given above, that theory does not have a vacuum with a mass gap\foot{
There is a subtlety here: the argument for absence of mass gap in the
$N=2$ theory depends on a chiral symmetry which is here explicitly
broken by coupling to the massive fields.  The breaking is, however,
irrelevant at low energies unless a twisted chiral superpotential
is generated.  This can be excluded by a variety of standard arguments.
   For instance, the twisted chiral superpotential
would have to be  independent of the mass of the massive
particles, as $\epsilon$ is chiral
rather than     twisted chiral. In  two-dimensional
superrenormalizable theories like
this one, the effects of massive particles on the massless   ones vanish
for large mass, except sometimes for ill-convergent 	one-loop
tadpoles (absent
here because as all fields are in the adjoint representation of $SU(n)$,
the tadpoles  vanish).
So the fact that the twisted chiral superpotential must be independent
of $\epsilon$  means that it cannot be generated at all. }
(and presumably breaks supersymmetry because of the inability to
screen the charge at infinity).
In case (ii), the low energy theory has at least one unbroken
$U(1)$ under which (by an argument given above) no component of the
charge at infinity is neutral, so supersymmetry is spontaneously
broken at the quantum level.

The interesting case therefore is case (iii).  In this case,
the gauge group is completely broken -- screening the charge
at infinity -- and all fields, including
$\phi$ and $\bar \phi$, get a mass.  This therefore is the desired unique
vacuum with mass gap corresponding to the $(m,n)$ string.

\newsec{Bound States Of Other $D$-Branes And $p$-Branes}

In this section, we briefly apply some of the same methods
to other questions involving bound states of $p$-branes.  For
any $p$, one has to look at the dimensional reduction of
ten-dimensional supersymmetric Yang-Mills theory to $p+1$
dimensions.
We first consider the Type IIB theory, and then more superficially
Type IIA.

\subsec{Type IIB}

\bigskip\noindent
{\it Instantons}

One might begin
with the Type IIB
 $-1$-branes or instantons.  They are given by the dimensional
reduction of supersymmetric Yang-Mills theory all the way to
zero dimensions.  Thus, in the $n$-instanton sector, all
ten space-time coordinates become  matrices $X^i,\,i=0,\dots ,9$,
while the fermion zero modes in the instanton sector will be
represented by the supersymmetric partners $\psi_\alpha,\, $
$\alpha=1,\dots,16$, also in the adjoint representation.
The instanton measure is an  integral over $X$ and $\psi$ weighted
by $e^{-I}$, with $I$ a multiple of the dimensionally-reduced
supersymmetric Yang-Mills action:
\eqn\ksnk{I=\sum_{i<j}\Tr [X^i,X^j]^2 +\sum_{i,\alpha,\beta}
\Gamma^{i}_{\alpha\beta}\Tr\psi^\alpha[X^i,\psi^\beta].}
(The $\Gamma$'s are gamma-matrices.)
This is the pure case of seeing the space-time coordinates
as non-commuting matrices.  For a supersymmetric minimum of the
action, of course, we minimize the action with $[X^i,X^j]=0$, and
then the matrices commute.  One can ask whether non-supersymmetric
instantons can be found as higher critical points of $I$ with
non-commuting $X$'s.  A scaling argument shows that there are
none: $I$ scales as $t^4$ under $X\to tX$, so any critical point
has $I=0$.

This description of the effective action for
$-1$-branes has an intriguing similarity to the ADHM description
of Yang-Mills instantons in four dimensions, which are determined
by matrices $X^i$, $ i=1,\dots,4$, obeying not $[X^i,X^j]=0$
but the self-dual counterpart
$[X^i,X^j]={1\over 2}\epsilon^{ijkl}[X_k,X_l]$.

\bigskip\noindent
{\it Three-branes}

We next move on to consider bound states of Dirichlet three-branes.
A bound state with the same world-volume structure as the
basic three-brane would correspond to a
 vacuum  with mass gap of $N=4$ $SU(n)$
supersymmetric Yang-Mills theory in four dimensions.  There are none,
since anomalous triangle diagrams of the
 $SU(4)$ $R$-symmetry currents imply -- by 't Hooft's old argument --
that this theory can have no vacuum with mass gap.
In fact, it is strongly believed that in this theory the moduli
space of vacua is exactly given by the classical
answer.   In three-brane language, this  means that
 the vacua are labeled by the positions
of the $n$ three-branes, up to permutation, so that there are no bound
states even with exotic world-volume structure.

One might think that this result on absence of three-brane bound states
is what is needed to justify the assumption in \strominger\ of
considering            in
conifold physics only simple, and not multiple,
wrapping of three-branes around collapsing three-cycles.  But this would
be too hasty a conclusion; we have argued here only the
absence of three-brane bound-states in flat space, and the conclusion
does not immediately carry over
to the Calabi-Yau context where the three-branes
are wrapped around a curved cycle.  The opposite
result in flat space would, however, have been undesireable; a bound state
in flat space would have led to a bound state in the conifold problem
when the radius of curvature is large enough (larger than the largest
length scale important in the structure of the flat space bound state),
and therefore, by
BPS-saturation, also when it is small.  By studying the soft
modes in $N=4$ super Yang-Mills theory on ${\bf R}\times {\bf S}^3$,
it can be seen that the bound state problem for $n $ three-branes
wrapped around an ${\bf S}^3$ is equivalent to the       question
-- which will not be addressed here --
of whether there are bound states at threshold in the dimensional
reduction to $0+1$ dimensions
of four-dimensional $N=1$ super Yang-Mills theory
with gauge group $SU(n)$.

\bigskip\noindent
{\it Five-Branes}

For Dirichlet five-branes the story is formally the same.
For bound states with the same structure as the elementary $D$-brane,
one must
consider vacua with mass gap
 in six-dimensional supersymmetric Yang-Mills theory.
There can be none because of anomalous four-point functions of the
$SU(2)\times SU(2)$ $R$-symmetry.  One might worry about the
unrenormalizability of six-dimensional super  Yang-Mills theory,
but this issue seems inessential as the string theory provides
some sort of cutoff, and the anomaly argument is an infrared argument.
The unrenormalizability -- and weak coupling in the infrared --
strongly suggest that the moduli space of vacua is given by the
classical answer, and hence that there are also no bound states
with exotic world-volume structure.

\nref\strom{A. Strominger, ``Heterotic Solitons,''
Nucl. Phys. {\bf B343} (1990) 167.}
\nref\cal{C. G. Callan, Jr., J. A. Harvey,
and A. Strominger, ``World-Sheet Approach To Heterotic
Instantons And Solitons,'' Nucl. Phys. {\bf B359} (1991) 611.}
\nref\dlk{M. J. Duff, R. R. Khuri, and J. X. Lu,
``String Solitons,'' Phys. Rept. {\bf 259} (1995) 213-326,1995,
hep-th/9412184. }
But the absence of     bound states of Dirichlet five-branes
with each other is not the whole story for five-branes.  One must
also consider the solitonic five-brane \refs{\strom - \dlk}.
In fact the two sorts of five-brane form an $SL(2,{\bf Z})$ doublet
just like the two kinds of string.  Electric-magnetic duality in ten
dimensions pairs the five-branes with strings.   $SL(2,{\bf Z})$ predicts
the existence of a bound state of $m$ Dirichlet five-branes with
$n$ solitonic ones for every relatively prime pair $m,n$.  The
prediction cannot be fully tested at the moment because the solitonic
five-branes are not sufficiently well understood, but a few
simple remarks are possible.

Note that in string units, the elementary string and $D$-string
have tensions of order $1$ and $1/\lambda$, respectively;
in Einstein units (where duality acts naturally) these become
$\sqrt\lambda$ and $1/\sqrt \lambda$. The
Dirichlet and solitonic five-brane have tensions of order
$1/\lambda$ and $1/\lambda^2$ , respectively, in string units;
in Einstein units the tensions are of order $\sqrt \lambda$ and
$1/\sqrt\lambda$.  There is thus a parallel between strings
and five-branes with the elementary string mapped to the Dirichlet
five-brane and the $D$-string mapped to the solitonic five-brane.
A BPS-saturated bound state of $m$ Dirichlet   five-branes and
$n$ solitonic ones would have tension in string units
\eqn\hins{\tilde T_{m,n}=
{\tilde T\over \lambda}\sqrt{m^2+{n^2\over \lambda^2}},}
a formula quite analogous to \ikko\ for strings.

The first bound state problem that we considered in section two was
a bound state of one $D$-string and $m$ elementary strings.  The
main qualitative issue was that the tension of the elementary string
completely disappears (in the weak coupling limit) in the presence
of of a $D$-string.  There is a similar issue for a bound state of
one solitonic five-brane with $m$ Dirichlet five-branes.  The $m$
Dirichlet five-branes in vacuum
would have a tension  of order $m/\lambda$,
but the above  formula says  that
for weak coupling this energy practically disappears in the field of
a solitonic five-brane.  Adding the $m$ Dirichlet five-branes to the
solitonic one should increase the ground state energy (in string
units) by an amount
only of order $m^2$, and not $m/\lambda$, for small $\lambda$.

How can this be?  The key is that the soliton five-brane is given
explicitly \refs{\strom - \dlk}
by a four-dimensional solution which contains a region
(an infinite tube that represents a sort of hole in space-time) in which
the dilaton blows up and the effective value of $\lambda$  goes to
infinity.  The energy of a Dirichlet five-brane therefore vanishes as
it falls down the hole.  To make this quantitative, one would have
to understand better the strong-coupling region of the soliton,
but one can at least assert that the main surprising feature of the
soliton solution, which is existence of the strong coupling end,
 is just what is needed  to make the $SL(2,{\bf Z})$
prediction possible.

\bigskip\noindent
{\it Seven-branes}

The Type  IIB theory also  has a Dirichlet seven-brane.  Its bound
states with standard world-volume structure
would correspond to vacua with mass gap in the dimensional
reduction of ten-dimensional super Yang-Mills theory to eight dimensions.
Such vacua do not exist, because the $U(1)$ global symmetry
(which arises in the dimensional reduction from ten to eight dimensions)
has an anomalous five point function, which would be impossible in
a vacuum with mass gap.  The weak infrared coupling of the theory
strongly suggests that bound states with exotic world-volume
structure are also absent.

\subsec{Type IIA Superstrings}

Now we move on to the Type IIA superstring in ten dimensions.
There are Dirichlet $p$-branes for $p$ even, while elementary
and solitonic $p$-branes only exist for odd $p$ (in fact, $p$ equal to
one or five).
So the only bound states to inquire about are the bound states of
$D$-branes with themselves, which correspond to
 vacua with mass gap in ten-dimensional $SU(n)$
super Yang-Mills theory dimensionally reduced to $p+1$ dimensions.

Predictions for small $p$  seem to follow from results
about string dynamics.  For $p=0$, where one is dealing with ordinary
particles (carrying Ramond-Ramond electric charge), precisely one
bound state for each $n$
is  apparently needed to agree with the Kaluza-Klein spectrum of
eleven-dimensional supergravity.
For $p=2$, to make sense of the physics of conifolds \strominger,
one wants no bound state when two-branes are wrapped around a
two-cycle, and therefore (taking the limit as the Calabi-Yau manifold
is scaled up) no bound state in flat space.  For $p>2$ there seem to be
no known predictions.

For $p=0$, one has ordinary quantum mechanics, albeit supersymmetric
quantum mechanics of a rather special sort.  One wants to know whether
there are bound states at threshold, and these might be accessible
to analysis, though the question is beyond the reach of the present paper.

For $p>0$, one is dealing with odd-dimensional quantum field theory,
and  anomaly-based arguments to exclude ground states with mass gap are
not nearly as powerful as they are in even dimesions.  In some cases,
however, some results can be obtained  using discrete anomalies.  For
instance, for $p=2$, the relevant three-dimensional
super Yang-Mills theory
has an $SO(7)$ global symmetry (obtained by dimensional reduction from
ten dimensions).  If one weakly gauges the $SO(7)$, then one can consider
whether the effective action of the theory is even or odd under a
topologically non-trivial $SO(7)$ gauge transformation.  For $n$ even
(so that the dimension of $SU(n)$ is odd), reasoning given  on p. 309
 of \ref\alv{L. ALvarez-Gaum\'e and E. Witten, ``Gravitational
Anomalies,'' Nucl. Phys. {\bf B234} (1983) 269.} shows that
 the effective action is odd, behavior that cannot be reproduced
in a parity-conserving theory with mass gap.  (The global anomaly
can be reproduced in a theory with mass gap
 by adding a parity-violating Chern-Simons interaction.)
So any  vacua with mass gap have
spontaneously broken parity and are paired by the action of parity; the
total number of bound states with standard world-volume structure
is therefore even.

\listrefs
\end

\\
Title: Bound States Of Strings And p-Branes
Author: Edward Witten
Comments: 25 pages, harvmac
\\
The recent discovery of an explicit conformal field theory
description of Type II $p$-branes makes it possible to
investigate the existence of bound states of such objects.
In particular, it is possible with reasonable  precision
to verify the prediction that the Type IIB superstring in ten dimensions
has a family of soliton and bound state
strings permuted by $SL(2,{\bf Z})$.  The space-time coordinates
enter tantalizingly in the formalism as non-commuting matrices.
\\
\end